\documentclass[11pt,a4paper]{article}
\usepackage{jheppub}
\usepackage{amsmath}
\usepackage{dsfont}
\usepackage{ulem}
\usepackage{natbib}
\usepackage{xcolor}
\usepackage[hang,flushmargin]{footmisc}
\usepackage{tikz-cd}

\makeatletter\renewcommand{\@biblabel}[1]{#1.}\makeatother

\def\be{\begin{equation}}
\def\ee{\end{equation}}
\def\W{{\rm W}}
\def\U{{\rm U}}
\def\i{{\rm i}}
\def\e{{\rm e}}
\def\d{{\rm d}}

\def\nn{{\nonumber}}

\def\ep{{\epsilon}}
\def\epp{{\epsilon_{\scriptscriptstyle +}}}

\newcommand{\mb}[1]{\mathbf{#1}}

\def\ket#1{{ |#1\rangle}}
\def\bra#1{{ \langle#1|}}
\def\braket #1#2{{ \langle #1|#2 \rangle}}

\def\HFN#1#2#3{ {}_N F_{N-1}{\tiny\left[\begin{array}{cc}#1\\ #2\end{array}|#3\right]}}

\def\s{{\mb s}}

\def\Y{{\mb Y}}
\def\y{{\mb y}}
\def\S{{\mb S}}
\def\Q{{\mb Q}}

\def\T{{\mb T}}

\def\Xint#1{\mathchoice
   {\XXint\displaystyle\textstyle{#1}}%
   {\XXint\textstyle\scriptstyle{#1}}%
   {\XXint\scriptstyle\scriptscriptstyle{#1}}%
   {\XXint\scriptscriptstyle\scriptscriptstyle{#1}}%
   \!\int}
\def\XXint#1#2#3{{\setbox0=\hbox{$#1{#2#3}{\int}$}
     \vcenter{\hbox{$#2#3$}}\kern-.5\wd0}}

\def\dashint{\Xint-}

\title{Quiver $\text{W}_{\ep_1,\ep_2}$ algebras of 4d $\mathcal{N}=2$ gauge theories}

\author[a]{Fabrizio Nieri}
\author[b]{and Yegor Zenkevich}

\affiliation[a]{DESY Theory Group, Notkestraße 85, 22607 Hamburg, Germany.}

\affiliation[b]{International School of Advanced Studies (SISSA), via Bonomea 265, 34136 Trieste, Italy\\[3pt]
  INFN, Sezione di Trieste,\\[3pt]
  ITEP, 25 Bolshaya Cheremushkinskaya street, 117218 Moscow, Russia\\[3pt]
  ITMP MSU, Leninskie gory 1, 119991 Moscow, Russia}

\emailAdd{fb.nieri@gmail.com}
\emailAdd{yegor.zenkevich@gmail.com}

\abstract{We construct an $\ep$-deformation of W algebras, corresponding to the additive version of quiver $\text{W}_{q,t^{-1}}$ algebras which feature prominently in the 5d version of the BPS/CFT correspondence and refined topological strings on toric Calabi-Yau's. This new type of algebras fill in the missing intermediate level between 
  $q$-deformed and ordinary W algebras. We  show
  that $\ep$-deformed W algebras are spectral duals of conventional
  W algebras, in particular the $\ep$-deformed conformal blocks manifestly reproduce instanton partition functions of 4d $\mathcal{N}=2$ quiver gauge theories in the full $\Omega$-background and give
  dual integral representations of ordinary W conformal blocks.}


\preprint{DESY 19-233~~ITEP-36/19}

\begin{document}

\maketitle
\flushbottom



\section{Introduction}
Any quantum field theory gives rise to an algebra of operators acting
on its Hilbert space. The virtue of supersymmetric gauge theories is
that for a certain BPS subsector of the Hilbert space, the algebra can
often be written explicitly. For 4d $\mathcal{N}=2$ quiver gauge
theories, the main focus of this paper, the space of supersymmetric
states is described by the (equivariant) cohomology of instanton
moduli spaces
\cite{Lossev:1997bz,Moore:1998et,Moore:1997dj,Losev:1997tp,Nekrasov:2002qd,Nekrasov:2003rj}. The
resulting algebra describes the action of certain correspondences on
the cohomology and is known as the spherical Hecke algebra or affine
Yangian
\cite{alex2014affine,maulik2012quantum,schiffmann2012cherednik}. Equivalently,
it can also be understood as the $\text{W}_{1+\infty}$ algebra
\cite{Prochazka:2015deb,Gaberdiel:2017dbk} of currents of all spins
from one to infinity (see e.g.~\cite{Awata:1994tf} for a review and
\cite{Gaberdiel:2010pz} for its appearance in AdS/CFT).
For a fixed representation, only the first few currents are
independent while the others can be expressed through them: these form
the $\text{W}$ algebra familiar from the 2d CFT context
\cite{Zamolodchikov:1985wn}. The most direct connection between 4d $\mathcal{N}=2$ gauge theories and $\text{W}$ algebras goes perhaps through the AGT
relation~\cite{Alday:2009aq, Wyllard:2009hg, Mironov:2009by}, namely
the identification of $\text{W}$ conformal blocks with partition
functions of class $\mathcal{S}$ theories~\cite{Gaiotto:2009we} in the
$\Omega$-background. This is a
remarkable corner of a more general framework known as the BPS/CFT
correspondence
\cite{2008arXiv0801,Nekrasov:2009uh,Nekrasov:2009ui,Nekrasov:2009rc,Nekrasov:2011bc,Carlsson:2013jka,Nekrasov:2015wsu,Nekrasov:2016qym,Nekrasov:2016ydq,Nekrasov:2017rqy,Nekrasov:2017gzb},
which allows the BPS sector of gauge theories with extended
supersymmetry to be studied through 2d CFT or integrability inspired
methods
\cite{Nekrasov:2009uh,Nekrasov:2009ui,Nekrasov:2009rc,Nekrasov:2011bc}. In
this paper we explore yet another corner, giving an orthogonal or
dual perspective on the AGT correspondence, at least when class
$\mathcal{S}$ theories and unitary quiver gauge theories overlap.

In fact, while the AGT philosophy relies on the description of class
$\mathcal{S}$ theories through Gaiotto's curve, our starting point is,
following~\cite{Nekrasov:2012xe,Nekrasov:2013xda}, closer to
Seiberg-Witten (SW) technology
\cite{Seiberg:1994rs,Seiberg:1994aj}. This change of perspective leads
to a double quantization of the SW geometry when the
$\Omega$-background is fully turned on~\cite{Kimura:2016ebq} and to
introduce a new type of algebras associated to 4d $\mathcal{N}=2$
quiver gauge theories which we call $\text{W}_{\ep_1,\ep_2}$
algebras. We show how this algebraic structure is
directly connected to instanton calculus and hence why it is useful
for computing partition functions of quiver gauge theories (possibly
coupled to defects). Moreover, we also explain how this type of
algebras can be thought as \textit{spectral dual} of ordinary
$\text{W}$ algebras, hence featuring joint conformal blocks in which
(roughly speaking) the notion of positions and conformal weights is
exchanged. This can be made more precise by studying certain surface
defects and related integrable systems.

\textbf{Spectral duality}. In order to understand the origin of such
duality, it is useful to recall that the AGT relation has a natural
uplift to 5d $\mathcal{N}=1$ theories on a circle
\cite{Awata:2009ur,Mironov:2011dk,Nieri:2013yra,Nieri:2013vba,Aganagic:2013tta,Aganagic:2014oia,Aganagic:2014kja}. In
this case, the partition functions of the 5d version of class
$\mathcal{S}$ theories~\cite{Benini:2009gi} are identified with the
($q$-deformed) conformal blocks of the $\text{W}_{q,t^{-1}}$ algebras
\cite{Shiraishi:1995rp,Awata:1995zk,FRqW}, a one-parameter deformation
of $\text{W}$ algebras to which they reduce upon taking what we call
the \textit{CFT-limit.} From the gauge theory perspective, this is
just one of the possible 4d limits, but a strong coupling one as it typically involves sending the compactification
radius to zero and the coupling constant to infinity. We would like to consider another 4d limit, which we
refer to as the \textit{$\hbar$-limit}, resulting in an effective
weakly coupled gauge theory description instead.

From the algebraic viewpoint, the $q$-deformed counterpart of the
affine Yangian is represented by the DIM algebra
\cite{Ding1997,miki2007}, which can also be viewed as a double quantum
deformation of the $\text{W}_{1+\infty}$ algebra. Notably, the DIM
algebra has a large group of automorphisms, including
$\text{SL}(2,\mathbb{Z})$. This group is identified with the type IIB
self-duality group~\cite{Bourgine:2018fjy}, and it is responsible for
highly non-trivial dualities between different 5d quiver gauge
theories~\cite{Bastian:2017ary} arising from geometric engineering \cite{Katz:1996fh,Hollowood:2003cv}. For instance, when applied to linear
quivers with unitary gauge groups all of the same rank, the action of
the S-element of $\text{SL}(2,\mathbb{Z})$ exchanges the rank of the
quiver with the rank of the gauge groups. In string/gauge theory, this
duality is commonly known as fiber/base or IIB S-duality
\cite{Katz:1997eq,Bao:2011rc}, while in the world of integrable
systems this is related to (bi)spectral or PQ-duality~\cite{adams1990dual,Harnad:1993hw,Bertola:2001hq,
  wilson1993bispectral}. We stick to
the latter term henceforth, which seems to be more universal. Understanding what is the remnant of the
spectral duality upon taking a 4d limit is the main goal of this
paper.

\textbf{5d perspective}. The $(2,0)$ little string theory (a deformation
of the 6d $\mathcal{N}=(2,0)$ theory away from the conformal point)
provides a powerful framework
\cite{Aganagic:2015cta,Haouzi:2016ohr,Haouzi:2017vec} for a deeper
understanding of the AGT correspondence. Before taking any 4d limit, an effective 5d gauge theory description is valid. However, it naively breaks down in the CFT-limit yielding a 4d class $\mathcal{S}$ theory
or $\text{W}$ algebra description. Importantly, the quiver gauge
theories which the little string naturally assigns to
$\text{W}_{q,t^{-1}}$ algebras are not the ones which the 5d lift of
AGT would assign, rather their spectral
duals~\cite{Zenkevich:2014lca,Morozov:2015xya}. This is particularly
evident by considering Kimura-Pestun (KP) construction
\cite{Kimura:2015rgi,Kimura:2017hez,Kimura:2019hnw} of
$\text{W}_{q,t^{-1}}(A_{M-1})$ free boson correlators with $N+2$
insertions, which capture the partition functions of
$\text{U}(N)^{M-1}$ theories rather than $\text{U}(M)^{N-1}$, the
latter being computed by $\text{W}_{q,t^{-1}}(A_{N-1})$ correlators
with $M+2$ points.\footnote{The so-called $\text{U}(1)$ factors play an
  important and subtle role in the gauge/algebra correspondence. In
  our setup we use almost exclusively $\text{U}(N)$ groups and not the
  $\text{SU}(N)$ ones.} From the purely gauge theoretic perspective, the
spectral duality may seem to be completely lost in the 4d limit
simply because in a dual pair of 5d theories, one becomes
strongly coupled as a 4d limit is taken. Similarly, on the algebraic
level it is not obvious whether the automorphisms of the DIM algebra
survive in the Yangian limit. Ultimately, the loss of manifest
$\text{SL}(2,\mathbb{Z})$ symmetry is one of the reasons why the AGT
relation is highly non-trivial.\footnote{It is worth noting that the
  action of the $\text{SL}(2,\mathbb{Z})$ automorphism group is already pretty involved even
  at the DIM algebra level. In particular, the automorphisms
  modify the Hopf algebra structure by a non-trivial Drinfeld
  twist, which encodes the choice of the natural
  basis in the
  CFT \cite{Alba:2010qc,Mironov:2013oaa,Zenkevich:2014lca,Bourgine:2018fjy}.}

\textbf{4d limits}. In this paper, rather than following the CFT
perspective when reducing from 5d to 4d, we favor the gauge theory
construction and consider a 4d limit which preserves the quiver description. In the setup where the little string geometry is
compactified on a cylinder with radius $\ell_c$ and punctures
represented by D5 defects wrapping vanishing 2-cycles in a singular transverse space, an effective description in terms of 5d quiver gauge
theories on the defects can be given. The resulting physics is weakly
coupled and five dimensional because the winding modes of the little
string manifest as KK modes on the dual cylinder of radius
$\hbar\equiv \ell_s^2/\ell_c$, with $\ell_s$ being the string length. The CFT-limit is indeed strongly coupled as upon sending
$\ell_s\to 0$, the modulus $\tau/\ell_s^2$ must be kept fixed, with
$\tau\equiv \ell_s^2/\ell_c g^2_5$ parametrizing the dimensionless
instanton chemical potential. However, we can consider another limit
in which the cylinder is decompactified while keeping $\ell_s$ finite:
this is what we have called the $\hbar$-limit, and it is indeed a 4d
weak coupling limit as the winding modes decouple while
$\tau\equiv1/g_4^2$ is fixed. Since $\hbar\to 0$ in both limits but either $\ell_s\to 0$ or $\ell_c\to\infty$, the CFT-limit and the $\hbar$-limit indeed seem to be related to dual descriptions of the
same physics. In the following, we study this possibility
from the vantage point of the algebraic interpretation of certain BPS
observables.

\begin{figure}[!ht]
\leavevmode
\begin{center}
\includegraphics[height=0.4\textheight]{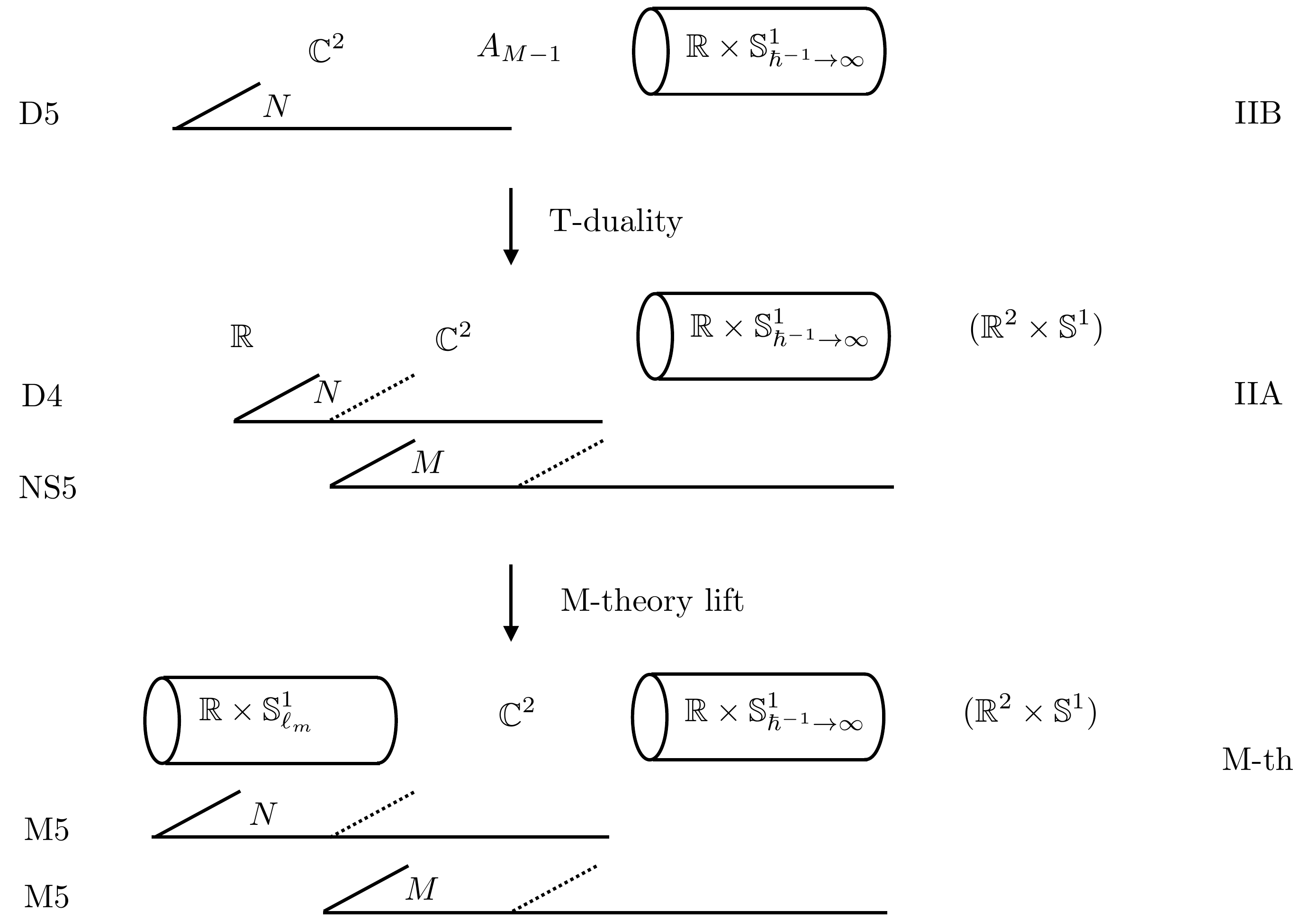}
\end{center}
\caption{Sketch of a type IIB geometry (top picture), its T-dual type
  IIA Hanany-Witten setup (middle picture) and the M-theory uplift
  (lower picture). The ``physical'' plane $\mathbb{C}^2$ on the left
  represents the space on which the gauge theory lives, $A_{M-1}$
  denotes the blowup of $\mathbb{C}^2/ \mathbb{Z}_M$ singularity in
  the transverse space, and the cylinder on the right is the internal
  space of the little string theory. T-duality along the circle fibers
  of the $A_{M-1}$ geometry produces $M$ additional NS5 branes
  spanning the ``physical'' $\mathbb{C}^2$. In the M-theory picture  there is an emerging cylinder. At a
  generic point on the Coulomb branch the M5 branes coalesce into a
  single brane wrapping the SW curve in the cotangent bundle of
  Gaiotto's base curve where a 2d CFT is supposed to live due to
  AGT. The symmetry between two cylinders in the M-theory setup gives
  rise to the spectral duality exchanging $M$ and $N$. Though the
  decompactification limit treats the cylinders differently (and hence
  the $\text{W}$ and $\text{W}_{\ep_1,\ep_2}$ algebras look different)
  the duality persists.}
\label{fig:Mcart}
\end{figure}

\textbf{Quiver algebras}. As expected, in the $\hbar$-limit the gauge
theory does not have a direct $\text{W}$ algebra description: rather,
the $\Omega$-background partition functions of 4d $\mathcal{N}=2$
quiver gauge theories with unitary groups can be manifestly identified
with ($\ep$-deformed) conformal blocks of dual algebras which
we call quiver $\text{W}_{\ep_1,\ep_2}$ algebras. The simplest example
where we know that the two complementary interpretations must coexist
is represented by the $\text{U}(N)$ SQCD: its $\Omega$-background
partition function is a $\text{W}_N$ conformal block with 2 full and 2
simple punctures (of the type featuring in the AGT duality) or a
$\text{W}_{\ep_1,\ep_2}(A_1)$ conformal block with $N+2$ points. In
fact, it might be more appropriate to say that $\text{W}$ and
$\text{W}_{\ep_1,\ep_2}$ algebras are spectral dual to each other, and
as such they exhibit joint conformal blocks. When a simple string
theory construction is available (as for this example), this may also
be argued by considering various brane pictures (Fig. \ref{fig:Mcart}) related by known
dualities.\footnote{As pointed out in
 ~\cite{Koroteev:2019byp}, an illuminating framework is actually
  provided by the gauge origami setup of~\cite{Nekrasov:2016ydq}.}
From this reasoning, the AGT correspondence can be seen as a
combination of an almost straightforward identification
($\text{W}_{\ep_1,\ep_2}$/gauge duality) with a highly non-trivial map
(spectral duality). Our construction of the quiver
$\text{W}_{\ep_1,\ep_2}$ algebras is inspired by the analogous KP
construction of quiver $\text{W}_{q,t^{-1}}$ algebras, and in fact it can be derived by carefully implementing the
$\hbar$-limit. On the algebraic side this limit is pretty non-trivial, and one
of our main results is a clear recipe how to do it in a constructive
way. Related results on the $\hbar$-limit of the algebraic engineering
technique in gauge theories have recently appeared in
\cite{Bourgine:2018uod} from a different angle.

\textbf{Defects and integrable systems}. Integrable systems provide a
powerful window into the BPS sector of gauge theories
\cite{Donagi:1995cf,Gorsky:1995zq,Gorsky:2000px}.  Spectral duality is
not a novelty in this
context~\cite{Mironov:2012uh,Mironov:2012ba,Mironov:2013xva}, and in
fact we can use it to gain a better understanding of the interplay
between $\text{W}/\text{W}_{\ep_1,\ep_2}$ algebras and 4d
$\mathcal{N}=2$ gauge theories too. It is important to make here a clear
distinction between two very different types of integrable
systems appearing in the gauge theory context: spectral duality
appears in both of them but the manifestations are different. The first type of systems are SW systems,
whose classical spectral curves coincide with the SW curves and the low energy physics is obtained from period integrals. For example, for 4d linear quiver gauge
theories these systems are XXX spin chains. The spectral duality
exchanges the two variables in the spectral curve equations of the SW
systems. In the $\Omega$-background, SW integrable systems and the
corresponding spectral curves get (doubly) quantized, so the two
variables defining the curve equation become non-commuting position and momentum operators. The spectral duality survives the
quantization~\cite{Mironov:2012uh,Mironov:2012ba,Mironov:2013xva}.

The second type of the integrable systems, on which we focus in this
paper, are easier to understand starting with $\Omega$-deformed
gauge theories from the very beginning. In this case, there is a family
of commuting quantum Hamiltonians acting on the Hilbert space of the
AGT dual 2d CFT. Similarly, there is another family of commuting
Hamiltonians acting on the Hilbert space featuring in KP construction. As we have discussed previously, the AGT and KP
setups are related by spectral duality, so the quantum integrable
systems should also be related to each other in this fashion. Indeed, the avatar of
the spectral duality for these systems is the so-called
PQ-duality. Classically, PQ-dual integrable systems share the same
phase space with an invertible symplectomorphism between the
action-angle variables of one system to angle-action variables of the
other~\cite{Ruijsenaars1988,Fock:1999ae}. At the quantum level, the
cleanest examples can perhaps be found in the many-body
Calogero-Moser-Sutherland (CMS) and Ruijsenaars-Schneider (RS) models:
considering the type of positions/momenta of the particles,
trigonometric/trigonometric or rational/rational models are self-dual,
while trigonometric/rational and rational/trigonometric are dual to
each other. Even though the Hamiltonians of non-self-dual systems are
very different looking, PQ-duality ensures that they share the same
eigenfunctions, hence determining a generalized Fourier
kernel. Interestingly enough, these systems also arise in our
context. For $A$-type algebras, it is known that
$\text{W}_{q,t^{-1}}(A_N)$ singular vectors are described by Macdonald
polynomials~\cite{Awata:1995zk}, which are tRS
eigenfunctions. Similarly, $\text{W}_N$ singular vectors are described
by Jack polynomials, which are tCMS eigenfunctions
\cite{Awata:1995np}. This can be seen as a consequence of the CFT
limit which takes $\text{W}_{q,t^{-1}}(A_N)$ to $\text{W}_N$, hence
tRS to tCMS and specializes Macdonald to Jack functions. On the other
hand, the $\hbar$-limit reduces tRS to rRS, which is PQ-dual to the
tCMS model. We show how the rRS eigenfunctions are naturally
constructed in terms of $\text{W}_{\ep_1,\ep_2}(A_N)$ screening
operators, hence supporting the claim that our algebras are spectral
dual to ordinary $\text{W}$ algebras.

From the gauge theory perspective, the rRS/tCMS eigenfunctions are
partition functions of two different 2d limits of the 3d self-mirror
$T[\text{U}(N)]$ theory
\cite{Gaiotto:2008ak,Bullimore:2014awa,Zenkevich:2017ylb,Aprile:2018oau}. In fact,
$\text{W}_{\ep_1,\ep_2}$ algebras also describe 2d $\mathcal{N}=(2,2)$
quiver gauge theories, which nicely reflects the possibility of
including codim-2 defects in the parent 4d $\mathcal{N}=2$
theories
\cite{Doroud:2012xw,Gomis:2014eya,Gukov:2014gja,Gomis:2016ljm}. This
relation between integrable systems and gauge theories also explains
why spectral duality acts very differently across dimensions:
$\text{W}_{q,t^{-1}}$ algebras of 5d or 3d theories are associated to
self-dual systems and hence spectral duality (cf. fiber/base or 3d
mirror symmetry) acts between weakly coupled gauge theories;
$\text{W}/\text{W}_{\ep_1,\ep_2}$ algebras of 4d or 2d theories are
associated to non-self-dual systems and hence spectral duality
(cf. S- or T-duality or 2d mirror symmetry) acts between theories
that may not be in the same class, such as a strong/weak or GLSM/LG
duality~\cite{Hori:2000kt,Aganagic:2001uw,Gomis:2012wy}.



The rest of the paper is structured as follows. In section
\ref{sec:5dKP}, we review KP construction of quiver
$\text{W}_{q,t^{-1}}$ algebras and their relation to 5d instanton and 3d vortex calculus. In section \ref{sec:hWalg}, we investigate the $\hbar$-limit of $\text{W}_{q,t^{-1}}$ algebras defining the
$\text{W}_{\ep_1,\ep_2}$ algebras. In particular, we give their
definition through the screening currents, bosonization formulas and
the corresponding conformal blocks which are explicitly equal to 4d
instanton or 2d vortex partition functions. In section \ref{sec:spectralD}, we
provide explicit comparisons with $\text{W}$ algebras and study the
PQ-duality of the relevant integrable systems. Finally, we present
our conclusions, further comments and open questions in section
\ref{sec:outlook}. The paper is supplemented by appendices, where
conventions, notations and some technical aspects of the computations
are collected.

\section{Quiver $\W_{q,t^{-1}}$ algebras, instantons and vortices}\label{sec:5dKP}

\begin{figure}[!ht]
\leavevmode
\begin{center}
\includegraphics[height=0.2\textheight]{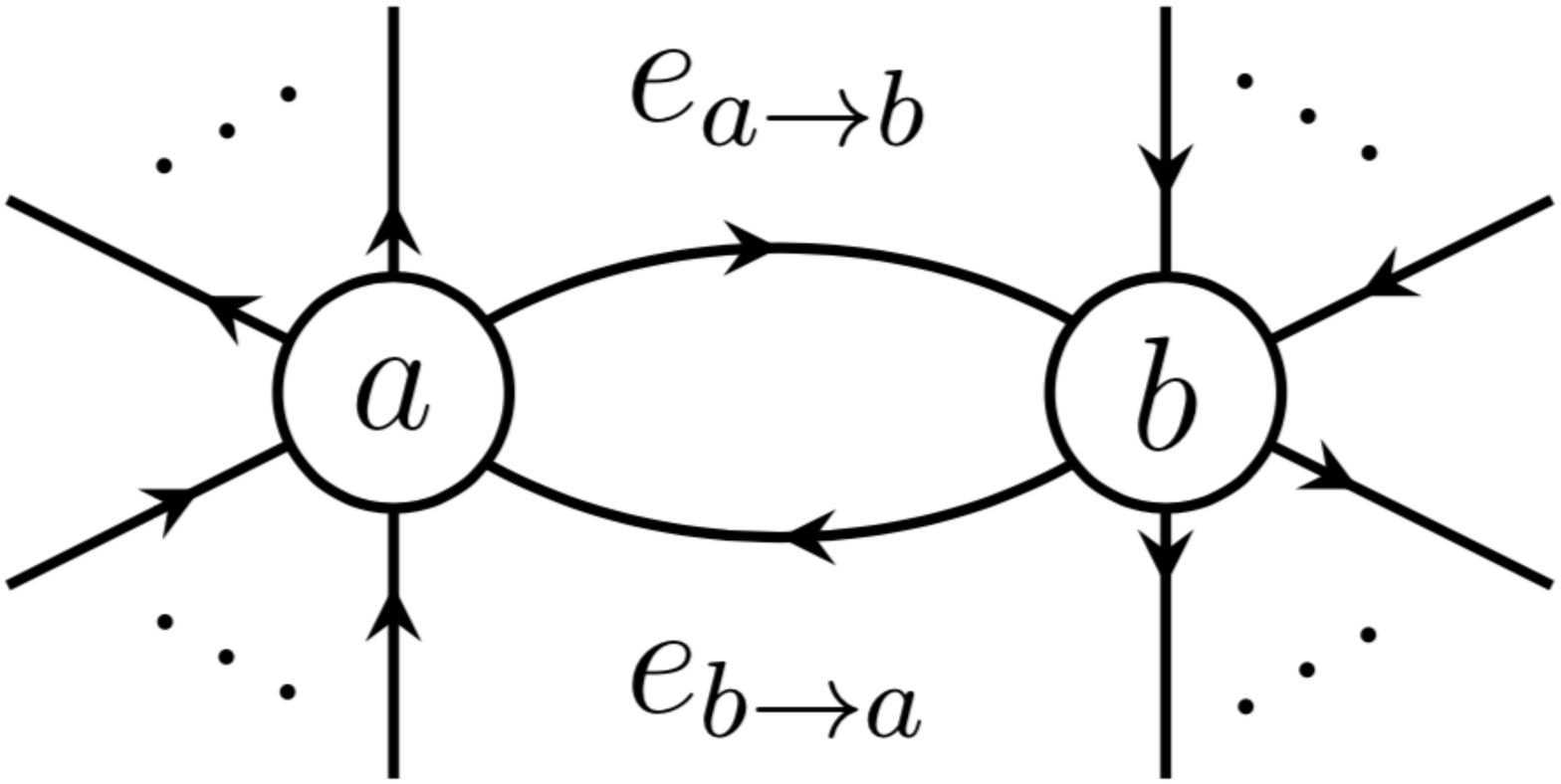}
\end{center}
\caption{Small piece of a quiver $\Xi$. Explicitly shown are two nodes $a,b\in\Xi_\circ$,  
arrows $e\in\Xi_\to$ from $a$ to $b$ or viceversa, and several arrows with source or target in $a$ or 
$b$.}
\label{Quiver}
\end{figure}

The main goal of this section is to review KP construction \cite{Kimura:2015rgi} of quiver 
$\W_{q,t^{-1}}$ algebras associated to 5d $\mathcal{N}=1$ Yang-Mills quiver gauge theories with unitary gauge groups, possibly coupled to 
(anti-)fundamental matter, in the 5d $\Omega$-background. For simplicity, we consider no Chern-Simons terms. The most important piece of common data is the quiver $
\Xi$, which is simply a collection of nodes $\Xi_\circ$ and arrows $\Xi_\to$ (see figure 
\ref{Quiver} for an example). The quiver determines the algebra $\W_{q,t^{-1}}(\Xi)$ and the interactions between 
gauge groups (nodes) through bi-fundamental matter (arrows). We assume that the quiver has no arrows without source or target since  this type of arrows, associated to (anti-)fundamental hypers, can be easily added later on. The structure of the quiver 
can compactly be encoded into its deformed Cartan matrix $\mathring{C}\in \mathbb{C}^{|\Xi_\circ|}
\times \mathbb{C}^{|\Xi_\circ|}$ (not necessarily associated to a Lie algebra). Given 
two nodes $a,b\in\Xi_\circ$ and arrows $e\in \Xi_{\scriptscriptstyle \to}$ between them, we define (slightly 
changing KP's definition)
\be\label{eq:qdefC}
\mathring{C}_{ab}\equiv (p^{1/2}+p^{-1/2})\delta_{ab}-\sum_{e:a\to 
b}p^{-1/2}\mu_e-\sum_{e:b\to a}p^{1/2}\mu_e^{-1}~,
\ee
where $p,\mu_e\in\mathbb{C}^\times$ are decoration
parameters associated to the nodes and arrows respectively, which are
part of the deformation parameters of the algebra, that is the
equivariant parameters of the gauge theory. The deformation parameters $q,t\in\mathbb{C}^\times$ are such that
$p=qt^{-1}$. In the gauge theory, they are usually parametrized by\footnote{Since the definitions make sense for complex parameters,  we omitted several factors of $\i$ w.r.t. other conventions in the existing literature.}
\be
q\equiv \e^{-\hbar\ep_1}~,\qquad t\equiv
\e^{\hbar\ep_2}~,
\ee
and are identified with the rotational equivariant parameters
of the 5d $\Omega$-background $\mathbb{C}^2_{q,t^{-1}}\times\mathbb{S}^1$, with $\hbar$ measuring the size of the $\mathbb{S}^1$ radius, while $\mu_e\equiv \e^{-\hbar M_e}$ represent
bi-fundamental (mass) fugacities.\footnote{In order to compare with
  \cite{Kimura:2015rgi}, we have to identify $(q)_{\rm
    here}=(q_2)_{\rm there}$, $(t)_{\rm here} =(q_1^{-1})_{\rm
    there}$. Also, the region of the parameter space may be restricted
  to ensure convergence of various expressions below.}

\subsection{The algebra}
For our purposes, the most convenient way to introduce the $\W_{q,t^{-1}}(\Xi)$ algebra is 
through its free boson realization. Let us start by defining a root-type Heisenberg 
algebra generated by oscillators $\{\s^a_n, n\in\mathbb{Z}_{\neq 0}\}_{a\in\Xi_\circ}$ 
and zero modes $\{\tilde \s^a_0,\s^a_0\}_{a\in\Xi_\circ}$ such that (we show non-trivial relations 
only)
\be\label{eq:scomm}
[\mb{s}^a_n,\mb{s}^b_m]=-\frac{1}{n}(q^{n/2}-q^{-n/2})(t^{-n/2}-t^{n/2})\mathring{C}^{[n]}_{ab}\ 
\delta_{n+m,0}\ ~,\quad [\tilde\s^a_0,\s^b_0]=\mathring{C}^{[0]}_{ab}~,
\ee
where the operation ${}^{[n]}$  replaces each parameter with its $n^{\rm th}$ 
power.\footnote{Note that our conventions are slightly different from those of \cite{FRqW}. e.g. our $\mb{s}$'s are related to their $\mb{a}$'s.} We next define the screening current
\be\label{eq:qdefS}
\S^a(x)\equiv \ :\e^{\s^a(x)}:~,\quad \s^a(x)\equiv -\sum_{n\neq 0}\frac{\s^a_n\ x^{-n}}{q^{n/2}-q^{-
n/2}} +\sqrt{\beta}\s^a_0+\sqrt{\beta}\tilde\s^a_0\ln x~,
\ee
where we have introduced $\beta\equiv \ln t/\ln q$ and the normal ordering symbol $:~:$ 
means moving all the non-positive modes to the left. The $\W_{q,t^{-1}}(\Xi)$ algebra 
can be defined as the commutant (up to total $q$-derivatives) of the screening 
current in the Heisenberg algebra, namely
\be\label{eq:screeningproperty}
\W_{q,t^{-1}}(\Xi)\equiv\{\T^a_n,n\in\mathbb{Z}\}_{a\in\Xi_\circ}\quad \text{such that}\quad  
[\T^a_{n},\S^b(x)]=\delta_{ab}\frac{\d_q}{\d_q x}\left(\cdots\right)~,
\ee
where we have defined the $q$-derivative
\be
\frac{\d_q}{\d_q x} f(x)\equiv \frac{f(q^{1/2} x)-f(q^{-1/2} x)}{(q^{1/2}-q^{-1/2})x}~.
\ee
A more direct definition of the algebra is also possible by introducing weight-type 
Heisenberg oscillators $\{\y^a_n, n\in\mathbb{Z}_{\neq 0}\}_{a\in\Xi_\circ}$ and zero 
modes $\{\tilde \y^a_0,\y^a_0\}_{a\in\Xi_\circ}$ satisfying
\be\label{eq:ycomm}
[\mb{y}^a_n,\mb{s}^b_m]=-\frac{1}{n}(q^{n/2}-q^{-n/2})(t^{-n/2}-t^{n/2})\delta_{ab}\ 
\delta_{n+m,0}\ ~,\quad [\tilde\y^a_0,\s^b_0]=[\tilde\s^a_0,\y^b_0]=\delta_{ab}~,
\ee
along with the fundamental vertices 
\be
\Y^a(x)\equiv \ :\e^{\y^a(x)}:~,\quad \y^a(x)\equiv \sum_{n\neq 0}\y^a_n x^{-n}+\sqrt{\beta}
\tilde\y^a_0\ln q+\ln p\sum_{b\in \Xi_\circ}(\mathring{C}^{-1})^{[0]}_{ab}~.
\ee
Then the $\W_{q,t^{-1}}(\Xi)$ generating currents $\T^a(x)\equiv \sum_{n\in\mathbb{Z}}
\T^a_n x^{-n}$  can be obtained by applying the iWeyl 
reflections  (the analogous of Weyl reflections for a 
root system) to the fundamental vertices\footnote{For the $A_M$ case, these generators were obtained from the quantum Miura transform \cite{Shiraishi:1995rp,Awata:1995zk}, while explicit expressions for simple Lie algebras were given in \cite{Frenkel:1997xxx}.}
\be\label{eq:Tcurr}
\T^a(x)\equiv \Y^a(p^{-1/2}x)+\textrm{all iWeyl reflections}~.
\ee
Because of this definition, the generating currents are also known as $qq$-characters (we refer to \cite{Lodin:2018lbz} for a recent gauge theory application, and \cite{Kim:2016qqs,Haouzi:2019jzk} for further gauge theory derivations in addition to the works by Nekrasov, Pestun and Shatashvili), a double quantization of SW geometries which also provide the deformation of $q$-characters \cite{frenkel1998qcharacters} of quantum affine algebras (see also \cite{KNIGHT1995187} for related work in the context of the Yangian algebras). Unfortunately, this type of expression rapidly becomes cumbersome as the number of nodes increases or if there are loops. However, for the simplest single node quiver with no arrows, i.e. $\Xi=A_1$, we get a sufficiently manageable expression for 
the generating current, namely
\be\label{eq:qVir}
\T(x)=\Y(p^{-1/2}x)+\Y(p^{1/2}x)^{-1}~.
\ee
This algebra is also known as the $q$-Virasoro algebra, and it can equivalently be defined 
as the associative algebra whose currents are subject to the quadratic relation
\be\label{eq:qVirdef}
\mathring{f}(w/z)\T(z)\T(w)-\mathring{f}(z/w)\T(w)\T(z)=-\frac{(1-q)(1-t^{-1})}{(1-p)}\left(\delta(p w/z)-\delta(pz/w)
\right)~,
\ee
where 
\be\label{eq:qVirf}
\mathring{f}(x)\equiv\sum_{n\geq 0}f_n x^n\equiv \exp\left(\sum_{n>0}\frac{(1-q^n)(1-t^{-n})}{(1+p^n)}\frac{x^n}{n}\right)= \frac{(p^2 x;p^2)_\infty (t^{-1} x;p^2)_\infty (q x;p^2)_\infty}{(x;p^2)_\infty(p t^{-1} 
x;p^2)_\infty(p q x;p^2)_\infty}
\ee
is the structure function and $\delta(x)\equiv \sum_{n\in\mathbb{Z}}x^n$ is the 
multiplicative $\delta$-function.

\subsection{Free field correlators}\label{subsec:Qinfinity}

The basic observables one can consider are free boson correlators between 
Fock states built on a (charged) Fock vacuum $\ket{\alpha}$ and its dual $\bra{\alpha}$, which are characterized by
\begin{align}
\s_{n>0}^a\ket{\alpha}&=0~,\quad \tilde\s_0^a\ket{\alpha}=\alpha^a\ket{\alpha}~,\quad \ket{\alpha}
\equiv\e^{\sum_{a}\alpha^a \y_0^a}\ket{0}~,\nn\\
\bra{\alpha}\s_{n<0}^a&=0~,\quad \bra{\alpha}\tilde \s_0^a=\alpha^a\bra{\alpha}~,\quad \bra{\alpha}
\equiv\bra{\alpha}\e^{-\sum_{a}\alpha^a \y_0^a}~,
\end{align}
with $\braket{0}{0}=1$. Because of the screening property (\ref{eq:screeningproperty}), correlators involving an arbitrary number of screening charges are of primary importance, e.g. to derive Ward identities or $q$-deformed conformal blocks in Dotsenko-Fateev representation \cite{Mironov:2011dk,Aganagic:2013tta,Aganagic:2014oia,Aganagic:2014kja}. There are two (complementary) definitions or routes one can pursue to define and use the screening charges, either by using standard contour integrals or Jackson integrals. In the first case, one simply defines the screening charges as
\be\label{def:ointS}
\Q^a\equiv \oint\d x\ \S^a(x)~,
\ee
where a choice of integration contour within correlators is to be understood, while in the second case one defines
\be\label{def:jackS}
\Q^a_z\equiv \int_z\d x_q\ \S^a(x)\equiv \sum_{k\in\mathbb{Z}} z q^k\ \S^a(z q^k)~.
\ee

\textbf{Remark}. In the latter definition, an arbitrary parameter $z$ has been introduced. This suggests to call $\Q^a_z$ the based screening charge with base point $z$. The possibility of introducing such parameter can be considered also in the contour integral representation, for instance by introducing a $q$-constant\footnote{These are functions $c_q(x)$ which in the multiplicative notation satisfy $c_q(qx)=c_q(x)$. They are closely related to elliptic functions.} which cannot be a priori fixed. Eventually, the two definitions can be made equivalent, but each one has its own virtues: as we are going to review below, the first definition is more useful for dealing with a finite number of screening charges or computing 3d vortex partition functions in the gauge theory side, while the second definition is more useful for dealing with an infinite number of screening charges or computing 5d instanton partition functions, where $z$ is mapped to the Coulomb branch parameter. The relation between the two approaches is detailed in \cite{Nieri:2017ntx}.    

Since one of the main goals of this section is to review KP construction of $\W_{q,t^{-1}}(\Xi)$ algebras and their relation to 5d instanton counting, we start by reviewing the second approach. Let us start by considering infinitely-many base points $z_{\emptyset^a_{Ai}}$ labeled by the nodes $a\in \Xi_\circ$ of the quiver $\Xi$ and two positive integers $(A,i)$. These points form the set $\chi_\emptyset \equiv \cup_{a\in\Xi_\circ}\chi^a_{\emptyset}$ (ground configuration) whose components are defined by\footnote{The appearing of indices of the type $Ai$ etc. stands for the pair $(A,i)$, i.e. no multiplication.}
\be
\chi^a_{\emptyset}\equiv \{z_{\emptyset^a_{Ai}} | A = 1, ..., N_a \ , i = 1,..., \infty\}~ .
\ee
Introducing auxiliary parameters $x_{aA}\in\mathbb{C}^\times$, we concretely set
\be
z_{\emptyset^a_{Ai}}\equiv x_{aA}t^{1-i}~.
\ee 
Similarly, let us also define the set $\chi \equiv \cup_{a\in\Xi_\circ}\chi^a$ in which each base point is shifted by an arbitrary integer power of $q$ determined by a sequence of integers $k^a_{Ai}\in\mathbb{Z}$ (excited configurations), namely
\be
\chi^a\equiv \{z_{\emptyset^a_{Ai}} q^{k^a_{Ai}}\ | A = 1, ..., N_a \ , i = 1,..., \infty\}~ .
\ee
As it will be clear in the next subsection, this set contains the fixed points of the torus action on the instanton moduli space arising from equivariant localization. In the following, it will be convenient to introduce the short-hand notation
\be
z_{k^a_{Ai}}\equiv z_{\emptyset^a_{Ai}} q^{k^a_{Ai}}=x_{aA}t^{1-i} q^{k^a_{Ai}}~.
\ee
We can now pick an order $\prec$ on $\chi$, which for definiteness we take as follows: $i$) we assign an order on $\Xi_\circ$ by simply choosing a labeling $a=1,\dots,|\Xi_\circ|$, and then we take $z_{k^a_{Ai}}\prec z_{k^b_{Bj}}$ if $a<b$; $ii)$ if $a=b$, then we take $z_{k^a_{Ai}}\prec z_{k^a_{Bj}}$ if $A<B$; $iii)$ if $A=B$, then we take $z_{k^a_{Ai}}\prec z_{k^a_{Aj}}$ if $i<j$. In fact, without loss of generality, we can take $|q|,|t^{-1}|<1$ and assume that $z_{k_{aAi}}\prec z_{k_{aBj}}$ corresponds to the radial ordering of the base points. Next, we define the operator  
\be\label{eq:Zoperator}
\widehat{\mb{Z}}\equiv  \prod^{\prec}_{z\in \chi_\emptyset}\Q^{\i(z)}_z=\sum_{\{k^a\}} \prod_{a=1}^{|\Xi_\circ|}\prod^{\prec}_{z\in \chi^a} z \ \S^{a}(z)  ~,
\ee
where $\i(z)$ simply denotes the index of the ground component to which $z$ belongs to, namely $\i(z)=a$ if $z\in\chi^a_\emptyset$. The crucial point here is the computation of the normal ordering function arising from commuting the non-positive modes of the screening currents to the right, which can be easily determined from the 2-point function
\be\label{eq:SaSb}
\S^a(x)\S^b(w)=\; :
\S^a(x)\S^b(w):\ \Delta^{a}_{\rm nodes}(w/x)^{\delta_{ab}} \ \Delta^{ab}_{\rm arrows}(w/x) \ x^{\beta C_{ab}^{[0]}}~,
\ee
where we have defined
\be
\Delta^{a}_{\rm nodes}(x)\equiv\frac{(x;q)_\infty (p x;q)_\infty}{(t x;q)_\infty(q x;q)_\infty}~,\quad \Delta^{ab}_{\text{arrows}}(x)\equiv\prod_{e:a\to b}\frac{(t \mu_e x;q)_\infty}{(\mu_e x;q)_\infty}\prod_{e:b\to a}\frac{(q x/\mu_e;q)_\infty }{(p  x/\mu_e;q)_\infty}~.
\ee
A straightforward computation yields
\begin{multline}
\prod_{a}\prod^{\prec}_{z\in \chi^a}  z\ \S^{a}(z)=\ :\prod_{a}\prod^{\prec}_{z\in \chi^a}  \S^{a}(z): \prod_a c^a_q(z_{\emptyset^a})\times\\
\times \prod_{(a,A,i)} z_{k^a_{Ai}}^{\sqrt{\beta}\widehat\alpha^a_\Xi}\prod_{a}\widehat\Delta^{a}_\text{ad.}(z_{k^a})\prod_{a<b}\widehat\Delta^{ab}_\text{bif.}(z_{k^a}/z_{k^b})~,
\end{multline}
where we have defined the functions
\begin{align}
\widehat{c}^a_{q}(z_{\emptyset^a})\equiv & \!\!\!\prod_{\substack{(A,i)\prec (B,j)}}\left(\frac{z_{\emptyset^a_{Ai}}}{z_{\emptyset^a_{Bj}}}\right)^{\frac{\beta}{2}\mathring{C}^{[0]}_{aa}}\frac{\Theta(t z_{\emptyset^a_{Ai}}/z_{\emptyset^a_{Bj}};q)}{\Theta(z_{\emptyset^a_{Ai}}/z_{\emptyset^a_{Bj}};q)}\prod_{e:a\to a}\frac{\Theta(\mu_e z_{\emptyset^a_{Ai}}/z_{\emptyset^a_{Bj}};q)}{\Theta(t \mu_e z_{\emptyset^a_{Ai}}/z_{\emptyset^a_{Bj}};q)}~,\\
\widehat\Delta^{a}_\text{ad.}(z_{k^a})\equiv & \!\!\!\prod_{\substack{(A,i)\neq (B,j)\\ A,B = 1,...,N_a\\ i,j = 1,...,\infty}}\!\!\!\frac{(x_{aA}/x_{aB}q^{k^a_{Ai}-k^a_{Bj}}t^{j-i};q)_\infty}{(t x_{aA}/x_{aB}q^{k^a_{Ai}-k^a_{Bj}}t^{j-i};q)_\infty}\prod_{e:a\to a}\frac{(t \mu_e x_{aA}/x_{aB}q^{k^a_{Ai}-k^a_{Bj}}t^{j-i};q)_\infty}{(\mu_e x_{aA}/x_{aB}q^{k^a_{Ai}-k^a_{Bj}}t^{j-i};q)_\infty}~,\\
\widehat\Delta^{ab}_\text{bif.}(z_{k^a}/z_{k^b})\equiv& \prod_{\substack{A = 1,...,N_a\\ B=1,\ldots, N_b\\ i,j = 1,...,\infty}}\prod_{e:a\to b}\frac{(t \mu_e x_{bB}/x_{aA}q^{k^b_{Bj}-k^a_{Ai}}t^{i-j};q)_\infty}{(\mu_e x_{bB}/x_{aA}q^{k^b_{Bj}-k^a_{Ai}}t^{i-j};q)_\infty}\times\nn\\
&\times \prod_{\substack{A = 1,...,N_a\\ B=1,\ldots, N_b\\ i,j = 1,...,\infty}}\prod_{e:b\to a}\frac{(t p\mu_e x_{aA}/x_{bB}q^{k^a_{Ai}-k^b_{Bj}}t^{j-i};q)_\infty}{(\mu_e p x_{aA}/x_{bB}q^{k^a_{Ai}-k^b_{Bj}}t^{j-i};q)_\infty}~,
\end{align}
and the (bare) charges 
\be
\widehat\alpha^a_{\Xi}\equiv \sqrt{\beta}\sum_{b}\frac{\mathring{C}^{[0]}_{ab}}{2}|\chi^b|-\frac{\mathring{C}^{[0]}_{aa}}{2}\sqrt{\beta}+\frac{1}{\sqrt{\beta}}~.
\ee
We have introduced the hats everywhere to remind us that we are considering infinitely-many variables, hence there can be subtleties. In fact, with an abuse of notation, we have denoted by $|\chi^a|$ the cardinality of the set $\chi^a$, which is of course infinite. We explain how to deal with these infinities in the following. Basically, all the divergences come from the zero modes of the screening currents, and this is due to the fact that we are considering infinite dimensional sets of insertion points. We can immediately see that  the simple factor
\be
\prod_{(a,A,i)} z_{k^a_{Ai}}^{\sqrt{\beta}\widehat\alpha^a_\Xi}=\prod_{(a,A,i)}\left(x_{aA}t^{1-i}\right)^{\sqrt{\beta}\widehat\alpha^a_\Xi}\times q^{\sum_{(a,A)} \sqrt{\beta}\widehat\alpha^a_\Xi|k^a_{A}|}~,
\ee
where we have set $|k^a_{A}|\equiv\sum_i k^a_{Ai}$, may be problematic. Note that the absolute value of $|k^a_A|$ can be large but finite for configurations with a finite number of finite parts $k^a_{Ai}$, namely $k^a_{Ai}=0$ for $i$ large enough. We will see momentarily how this comes about. The first factor of the expression above is ill-defined but this type of divergence is independent of the excited configuration and normalized observables would be not affected. Similarly, the function $c^a_q$ is invariant w.r.t. shifts of its arguments by integer powers of $q$, so it can also be pulled out of the summation over the excited configurations and neglected for the type of computations we are considering. Therefore, the only term needing our attention is the last one of the expression above because of the momentum $\widehat\alpha_\Xi$, which is infinite because of the finite amount of momentum carried by each of the infinite number of screening charges. It can be dealt with by considering a simple renormalization-like procedure, namely by letting $\widehat{\mb{Z}}$ to act on a charged Fock vacuum $\ket{\widehat\alpha_0}\equiv \ket{\widehat\alpha_\text{ren.}-\widehat\alpha_\Xi}$ capable of absorbing the infinite amount of momentum, with $\widehat\alpha_\text{ren.}$ finite.

\subsection{5d and 3d quiver gauge theories}\label{subsec:5dGT}

\subsubsection*{5d $\mathcal{N}=1$ theories: infinitely-many screening charges}

As the notation suggests, the various  $\widehat\Delta$ functions have a direct gauge theory interpretation. Indeed, they are essentially the 5d (K-theoretic) Nekrasov functions of the various building blocks defining the quiver gauge theory. The nodes correspond to unitary gauge groups $\U(N_a)$ with Yang-Mills couplings encoded by $\widehat\alpha^a_{\text{ren.}}$, the oriented arrows between the nodes to bi-fundamental or adjoint matter with mass parameters encoded in $\mu_e$,  the $x_{aA}$ to Coulomb branch parameters and $q,t$ to the $\Omega$-background equivariant parameters. This correspondence can be made explicit and precise by recalling the definition of the 5d Nekrasov function for an arbitrary pair of partitions $Y,W$ (see e.g. \cite{Awata:2008ed}) 
\begin{align}\label{eq:5dNek}
N_{Y W}(x)\equiv&\prod_{(i,j)\in Y}(1-x q^{Y_i-j}t^{W^\vee_j-i+1})
\prod_{(i,j)\in W}(1-x q^{-W_i+j-1}t^{-Y^\vee_j+i})=\\
=&\prod_{\substack{i = 1,...,\infty \\ j = 1,...,\infty}}\frac{(t x q^{Y_{i}-W_{j}}t^{j-i};q)_\infty}{(x q^{Y_{i}-W_{j}}t^{j-i};q)_\infty}\frac{(x t^{j-i};q)_\infty}{(t xt^{j-i};q)_\infty}~,
\end{align}
where $Y_i$ denotes the number of boxes in the $i^\text{th}$ row of $Y$ and $Y^\vee$ denotes the transpose diagram.  However, notice that for fixed $(a,A)$ the configuration of integers $\{k^a_A\}$ that we have considered so far does not generically correspond to a partition. In fact, no ordering $k^a_{Ai}\leq k^a_{A,i+1}$ nor positivity was assumed. However, it turns out that the normal ordering function of the nodes vanishes identically if the configuration $\{k^a_A\}$ is not a partition. Due to this observation, the sum over $\{k^a\}$ in the operator (\ref{eq:Zoperator}) is effectively a sum over colored partitions, and the state
\be
\ket{\widehat{\mb{Z}}}\equiv \widehat{\mb{Z}}\ket{\widehat\alpha_0}
\ee
corresponds to the (unnormalized) time-extended instanton partition function  of the quiver gauge theory. The time-extension refers to the fact that, as long as we do not project this state with a dual state, $\ket{\widehat{\mb{Z}}}$ still depends on the negative oscillators which can be identified with additional (higher-time) parameters \cite{Losev:2003py,Marshakov:2006ii} using the representation 
\be\label{def:timerep}
\frac{\s^a_{-n}}{q^{n/2}-q^{-n/2}}\simeq \tau^a_{n}~,\qquad \frac{n \,\y^a_{n}}{t^{n/2}-t^{-n/2}}\simeq \partial_{\tau^a_n}~,\quad n>0~.
\ee
In matrix model terminology, these parametrize the shape of the potential. Upon setting $\tau^a_n=0$ or (equivalently) projecting with a dual Fock state, we get the 5d instanton partition function of the theory in the usual form. For instance, for the simplest $\Xi=A_M$ quiver we have  the (normalized) result
\begin{multline}
\frac{\braket{\widehat\alpha_\infty}{\widehat{\mb{Z}}}}{\bra{\widehat\alpha_\infty}\prod_{a=1}^{M}\prod^{\prec}_{z\in \chi^a_\emptyset}  z\ \S^{a}(z)\ket{\widehat\alpha_0}}=\sum_{\{k^a\}}\frac{\prod_{a=1}^{M}\mathring{\Lambda}_a^{\sum_{A=1}^{N_a}|k^a_A|}}{\prod_{\substack{a=1,\ldots,M\\ 1\leq A, B\leq N_a}}N_{k^a_A k^{a}_B}(x_{aA}/x_{aB})}\times\\
\times\prod_{\substack{a=1,\ldots,M-1\\1\leq A \leq N_a\\ 1\leq B\leq N_{a+1}}}N_{k^a_A k^{a+1}_B}(p\mu_{a,a+1}x_{aA}/x_{a+1,B})\equiv Z^{\mathbb{C}^2\times\mathbb{S}^1}_\text{inst.}[A_M]~,
\end{multline}
where we have denoted the bi-fundamental fugacities by $\mu_{a,a+1}\equiv\mu_{e:a+1\to a}$ and the instanton counting parameters by $\mathring{\Lambda}_a\equiv q^{\sqrt{\beta}\widehat\alpha_{\text{ren.}}^a}$, while $\widehat\alpha_\infty$ is chosen to guarantee charge conservation. Note that the coupling to (anti-)fundamental matter, which can be associated to the presence of additional arrows with either no target or source, was not considered so far. It can be included by giving a background to the time variables of the form 
\be
\tau_n^a\to \tau^a_n-\sum_f\frac{\mu_{af}^n}{n(1-q^n)(1-t^{-n})}~,
\ee
with $\mu_{af}$ encoding the fundamental mass parameters, or (more generally) by the insertion of additional vertex operators
\be
\mb{V}^a(\mu)\equiv \ :\e^{\mb{v}^a(\mu)}:~,\quad \mb{v}^a(\mu)\equiv \sum_{n\neq 0}\frac{\mb{y}^a_n\, \mu_a^{-n}}{(q^{n/2}-q^{-n/2})(t^{-n/2}-t^{n/2})}~,
\ee
whose effect is inserting the appropriate $N_{\emptyset k^a_A}(x_{aA}/\mu)$ or $N_{k^a_A\emptyset}(\bar\mu/x_{aA})$ factors in the instanton partition function computed above.

\subsubsection*{3d $\mathcal{N}=2$ theories.}\label{subsec:3dGT}

In case we would like to use the more conventional definition (\ref{def:ointS}) of the screening charges, the analogous of the $\widehat{\mb{Z}}$ operator is 
\be
\mb{Z}\equiv \oint\prod_{a=1}^{|\Xi_\circ|}\prod_{i=1}^{r_a}\frac{\d x_{ai}}{2\pi\i x_{ai}}\ x_{ai}\ \S^a(x_{ai})~,
\ee
where we have considered a finite number $r_a$ of insertions for each type. Then we can write
\begin{multline}
\prod_{a=1}^{|\Xi_\circ|}\prod_{i=1}^{r_a} x_{ai}\ \S^a(x_{ai})=\ :\prod_{a=1}^{|\Xi_\circ|}\prod_{i=1}^{r_a} \S^a(x_{ai}): \prod_a c^a_q(x_a)\times\\
\times \prod_{a}\prod_{i=1}^{r_a} x_{ai}^{\sqrt{\beta}\alpha^a_\Xi}\ \prod_{a}\Upsilon^a_\text{ad.}(x_a)\prod_{a<b}\Upsilon^{ab}_\text{bif.}(x_a/x_b)~,
\end{multline}
where we have defined the functions
\begin{align}
\Upsilon^a_\text{ad.}(x_a)\equiv &\prod_{\substack{1\leq i\neq j\leq r_a}}\frac{(x_{ai}/x_{aj};q)_\infty}{(t x_{ai}/x_{aj};q)_\infty}\prod_{e:a\to a}\frac{(t\mu_e x_{ai}/x_{aj};q)_\infty}{(\mu_e x_{ai}/x_{aj};q)_\infty}~,\nn\\
\Upsilon^{ab}_\text{bif.}(x_{a}/x_{b})\equiv& \prod_{\substack{i = 1,...,r_a\\ j=1,\ldots, r_b}}\prod_{e:a\to b}\frac{(t \mu_e x_{bj}/x_{ai};q)_\infty}{(\mu_e x_{bj}/x_{ai};q)_\infty}\prod_{e:b\to a}\frac{(t p\mu_e^{-1} x_{bj}/x_{ai};q)_\infty}{(\mu_e^{-1} p x_{bj}/x_{ai};q)_\infty}~,\nn\\
c^a_{q}(x_a)\equiv& \prod_{\substack{i<j}}\left(\frac{x_{ai}}{x_{aj}}\right)^{\frac{\beta}{2}C^{[0]}_{aa}}\frac{\Theta(t x_{ai}/x_{aj};q)}{\Theta(x_{ai}/x_{aj};q)}\prod_{e:a\to a}\frac{\Theta(\mu_e x_{ai}/x_{aj};q)}{\Theta(t \mu_e x_{ai}/x_{aj};q)}~,\label{eq:qdefDeltas}
\end{align}
and the momentum 
\be
\alpha^a_{\Xi}\equiv \sqrt{\beta}\sum_{b}\mathring{C}^{[0]}_{ab}r_b-\frac{\mathring{C}^{[0]}_{aa}}{2}\sqrt{\beta}+\frac{1}{\sqrt{\beta}}~.
\ee

As the names suggest, these pieces can immediately be recognized as the 1-loop determinants of various 3d $\mathcal{N}=2^*$ multiplets on $\mathbb{C}_q\times\mathbb{S}^1$ as computed by supersymmetric localization (see e.g. \cite{Yoshida:2014ssa}).\footnote{Even though we occasionally use the $\mathcal{N}=4$ language, the localization results hold more generally for $\mathcal{N}=2$ preserving couplings. Hence, an hyper should be really though of as a pair of chirals in opposite gauge representations etc.. In particular, the mass parameters may be all independent.} The associated gauge theory has unitary gauge groups $\U(r_a)$ corresponding to the nodes, gauge holonomies in the Cartan torus parametrized by the insertion points $x_{ai}$, bi-fundamental (or adjoint) hypers with fugacities $\mu_e$ corresponding to the oriented arrows between the nodes and FI parameters encoded by the total momentum. The parameter $q$ is identified with the disk equivariant parameter while $t$ with the fugacity of the adjoint chiral. Therefore, using the representation (\ref{def:timerep}), the state
\be
\ket{\mb{Z}}\equiv \mb{Z}\ket{\alpha_0}
\ee
can be identified with the time-extended partition function of the quiver gauge theory \cite{Nedelin:2016gwu}. Upon setting the time variables to zero or (equivalently) projecting with a dual Fock state, we get the 3d holomorphic block in the (almost\footnote{This refers to the presence of the $c^a_q(x_a)$ functions, see the Remark below.}) canonical form \cite{Beem:2012mb}, namely
\be\label{eq:Z3d}
Z^{\mathbb{C}\times\mathbb{S}^1}[\Xi]\propto \braket{\alpha_\infty}{\mb{Z}}~,
\ee
\be
\braket{\alpha_\infty}{\mb{Z}}= \oint \prod_a \prod_{i=1}^{r_a} \frac{\d x_{ai}}{2\pi\i x_{ai}} \ \prod_a c^a_q(x_a)
\ \prod_{a}\prod_{i=1}^{r_a} x_{ai}^{\zeta^a}\ \prod_{a}\Upsilon^a_\text{ad.}(x_a)\prod_{a<b}\Upsilon^{ab}_\text{bif.}(x_a/x_b)~,
\ee
where we have identified the FI parameters $\zeta^a\equiv \alpha^a_0+\sqrt{\beta}\alpha^a_\Xi$ and $\alpha_\infty$ guarantees charge conservation. Note that the information about the base points or Coulomb parameters  of the previous subsection is apparently lost, simply because the definition of the screening charges does not involve them in the first place. However, in this case an integration contour needs to be specified. From the gauge theory perspective, this can be done more easily when the theory admits a discrete series of massive Higgs vacua, such as when coupling to  (anti-)fundamental matter. Such coupling  can be implemented by shifting the time variables according to 
\be
\tau^a_n\to \tau^a_n+\sum_{f=1}^{N_\text{f}^a}\frac{\mu_{af}^n-\bar\mu_{af}^{n}}{n(1-q^n)}~,
\ee
or by the insertion of additional vertex operators
\be
\mb{H}^a(\mu|\bar\mu)\equiv \ :\mb{V}^a(q^{1/2}/\mu)\mb{V}^a(q^{1/2}/\bar \mu)^{-1}:~,
\ee
with the effect of including factors of
\be
\Upsilon^a_\text{f}(x_a)\equiv \prod_{i=1}^{r_a}\prod_f \frac{(\bar \mu_{af} x_{ai};q)_\infty}{(\mu_{af}x_{ai};q)_\infty}
\ee
under the integral. The contour integral is then defined by a particular distribution of the integration variables $x_{ai}$ around distinct poles, namely by a partition of $r_a$ into $N^a_\text{f}$ parts of $r_{af}$ each. These additional parameters (a.k.a. the filling fractions) are known to be related to the Coulomb branch parameters \cite{Aganagic:2013tta,Aganagic:2014oia}. This can be explicitly checked in selected examples (e.g. SQCD) when Higgsing a parent 5d theory yields an effective 3d description, in which case it is easy to show the collapsing of the instanton partition function to its vortex counterpart \cite{Dimofte:2010tz,Bonelli:2011fq,Fujimori:2015zaa} (i.e. the non-perturbative contribution to the holomorphic blocks) upon discrete choices of $x_{aA}$ parametrized by positive integers $r_{af}$ (in fact, the cut-off on the number of rows of the Young diagrams $k^a$). Alternatively, the Coulomb branch parameters can also be related to additional multiplets/vertex operators as in the interpolation formulas found in \cite{Nieri:2017ntx}.

\textbf{Remark}. The function $c_q^a$ can presumably be thought of as the contribution of boundary degrees of freedom on the torus $\partial (\mathbb{C}\times\mathbb{S}^1)\simeq \mathbb{T}^2$ \cite{Benini:2013nda,Benini:2013xpa,Gadde:2013ftv}. The precise identification and the origin of the boundary theory, though interesting in its own right, is beyond the scope of this paper. In our discussion it does not play a significative role since once (anti-)fundamentals/vertex operators are included, the holomorphic blocks/correlators are defined via residues series with the relevant poles differing by integer powers of $q$. Hence the contribution of $q$-constants can ultimately be pulled out of integration and neglected.

\subsection{4d and 2d quiver gauge theories}\label{sec:4d2dgauge}\label{subsec:gaugelimit}

We now turn to looking at the theories of our main interest, namely the simple dimensional reduction of the quiver gauge theories of the previous subsections. Starting from the 5d setup, we are led to consider 4d $\mathcal{N}=2$ quiver gauge theories with unitary gauge groups,\footnote{The trace part can usually be stripped off from the computations by hand if needed, see e.g. \cite{Alday:2009aq}.} possibly coupled to 
(anti-)fundamental matter, in the 4d $\Omega$-background $\mathbb{C}^2_{\ep_1,\ep_2}$. The most important piece of data is again the very same quiver $\Xi$, to which we can associated a deformed Cartan matrix, now conveniently presented in additive notation as
\be\label{eq:4dcartan}
C_{ab}\equiv 2\cosh(\epp/2)\delta_{ab}-\sum_{e:a\to b}\exp(\ep_+/2-m_e)-\sum_{e:b\to a}\exp(-\ep_+/2+m_e)~,
\ee
where we have defined $\epp\equiv \ep_1+\ep_2$. Each node $a\in\Xi_\circ$ is associated to a $\text{U}(N_a)$ gauge group and each oriented arrow $\Xi_{\scriptscriptstyle\to}\ni e:a\to b$ to a bi-fundamental (or adjoint if $a=b$) hyper with mass $m_e$. For the time being, this matrix is just going to be a compact gadget to encode the relevant gauge theory data.

The basic building block for writing down the instanton partition
function of the quiver gauge theory is the 4d (cohomological) Nekrasov
function (see e.g. \cite{Nekrasov:2003rj,Sulkowski:2009ne})
\begin{align}\label{eq:4dNek}
n_{Y W}(X)\equiv&\prod_{(i,j)\in Y}(X +\ep_1(Y_i-j)-\ep_2(W_j^\vee-i+1))\times\nn\\
&~~~~~~~~~~~~~~~~\times\prod_{(i,j)\in W}(X+\ep_1( -W_i+j-1)-\ep_2(-Y^\vee_j+i))=\\
=&\prod_{\substack{i = 1,...,\infty \\ j = 1,...,\infty}}\frac{\Gamma_1(X +\ep_1(Y_{i}-W_{j})-\ep_2(j-i)|\ep_1)}{\Gamma_1(-\ep_2+X+\ep_1( Y_{i}-W_{j})-\ep_2(j-i)|\ep_1)}\frac{\Gamma_1(-\ep_2+X -\ep_2(j-i)|\ep_1)}{\Gamma_1(X -\ep_2 (j-i)|\ep_1)}~,
\end{align}
where the $\Gamma_1$ function is simply related to the Euler Gamma function by (appendix \ref{sec:appFun})
\be
\Gamma_1(X|\ep_1)\equiv \frac{\e^{(X/\ep_1-1/2)\ln\ep_1}}{\sqrt{2\pi}}\ \Gamma(X/\ep_1)~.
\ee 
For instance, the instanton partition function of the linear $\Xi=A_M$ quiver reads
\begin{multline}\label{eq:4dAM}
Z^{\mathbb{C}^2}_\text{inst.}[A_M]\equiv\sum_{\{k^a\}}\frac{\prod_{a=1}^{M}\Lambda_a^{\sum_{A=1}^{N_a}|k^a_A|}}{\prod_{\substack{a=1,\ldots,M\\ 1\leq A, B\leq N_a}}n_{k^a_A k^{a}_B}(X_{aA}-X_{aB})}\times\\
\times\prod_{\substack{a=1,\ldots,M-1\\1\leq A \leq N_a\\ 1\leq B\leq N_{a+1}}} n_{k^a_A k^{a+1}_B}(\epp +m_{a,a+1}+X_{a+1,B}-X_{aA})~,
\end{multline}
where $X_{aA}$ are the Coulomb branch parameters, $m_{a,a+1}\equiv m_{e:a+1\to a}$ the bi-fundamental masses and $\Lambda_a$ the instanton counting parameters. (Anti-)fundamental matter, which can be associated to arrows with either no source or target, can be easily coupled to the nodes by considering the appropriate $n_{\emptyset k^a_A}(-X_{aA}+m_f)$ or $n_{k^a_A\emptyset}(X_{aA}-\bar m_f)$ insertions.

Similarly, we can also consider codimension 2 theories supported on $\mathbb{C}_{\ep_1}$ and associated to the very same quiver $\Xi$. In 2d $\mathcal{N}=(2,2)$ language, we associate each node $a\in\Xi_\circ$ to a $\text{U}(r_a)$ vector multiplet coupled to an adjoint chiral of mass $-\ep_2$, and to each arrow $\Xi_{\scriptscriptstyle\to}\ni e:a\to b$ between the nodes we associate a pair of bi-fundamental chirals of mass $m_e$. The disk partition function can be computed by localization  \cite{Hori:2013ika,Honda:2013uca} and it reads
\be\label{eq:Z2d}
Z^{\mathbb{C}}[\Xi]\equiv\oint \prod_a \prod_{i=1}^{r_a} \frac{\d \sigma_{ai}}{2\pi\i} \ 
 \prod_{a}\prod_{i=1}^{r_a} \e^{\xi^a\sigma_{ai}}\ \prod_{a}\gamma^a_\text{ad.}(\sigma_a)\prod_{a<b}\gamma^{ab}_\text{bif.}(\sigma_a-\sigma_b)\prod_a \gamma^a_\text{f}(\sigma_a)~,
\ee
where $\sigma_a$ denotes the constant value of the vector multiplet scalar in the Cartan subalgebra of the gauge group, $\xi$ the FI parameters while the $\gamma$ functions represent the 1-loop determinants of the various multiplets. In particular, the vector and adjoint contribution associated to each node is
\be
\gamma^a_\text{ad.}(\sigma_a)\equiv \prod_{\substack{1\leq i\neq j\leq r_a}}\frac{\Gamma_1(-\ep_2+\sigma_{ai}-\sigma_{aj}|\ep_1)}{\Gamma_1(\sigma_{ai}-\sigma_{aj}|\ep_1)}\prod_{e:a\to a}\frac{\Gamma_1(m_e+\sigma_{ai}-\sigma_{aj}|\ep_1)}{\Gamma_1(\bar m_e+\sigma_{ai}-\sigma_{aj}|\ep_1)}~,
\ee
the total contribution of the pairs of bi-fundamental chirals between two nodes is
\be
\gamma^{ab}_\text{bif.}(\sigma_a-\sigma_b)\equiv  \!\!\!\!\! \prod_{\substack{i = 1,...,r_a\\ j=1,\ldots, r_b}}\prod_{e:a\to b}\frac{\Gamma_1(\sigma_{bj}-\sigma_{ai}-m_e|\ep_1)}{\Gamma_1(\sigma_{bj}-\sigma_{ai}-\bar m_e|\ep_1)}\prod_{e:b\to a}\frac{\Gamma_1(\sigma_{ai}-\sigma_{bj}+m_e|\ep_1)}{\Gamma_1(\sigma_{ai}-\sigma_{bj}+\bar m_e|\ep_1)}~,
\ee
while the possible contribution from pairs of fundamental/anti-fundamental chirals of (complexified) mass $m_f/\bar m_f$ associated to the arrows with either no source or target has been denoted by
\be
\gamma^{a}_\text{f}(\sigma_a)\equiv \prod_{i=1}^{r_a}\prod_{f}\frac{\Gamma_1(\sigma_{ai}+m_{af}|\ep_1)}{\Gamma_1(\sigma_{ai}+\bar m_{af}|\ep_1)}~.
\ee

\subsection{The 5d/3d~$\to$~4d/2d $\hbar$-limit}

The partition functions of the 4d and 2d quiver gauge theories can also be obtained from their 5d and 3d counterparts described in subsection \ref{subsec:5dGT} through a limiting procedure shrinking the size of the compactification circle. In fact, since all the 5d/3d variables (either gauge of global fugacities) can be interpreted as (complexified) holonomies along the circle, we can take the 4d/2d  limit we are interested in by setting the 5d/3d fugacities to be exponentials of the would-be 4d/2d variables multiplied by the $\mathbb{S}^1$ radius $\hbar>0$ and letting the latter go to zero, schematically
\be
(\cdots)_\text{5d/3d}\equiv \e^{-\hbar(\cdots)_\text{4d/2d}}~.
\ee 
In particular, we recall the identification 
\be
q\equiv \e^{-\hbar \ep_1}~,\quad t\equiv \e^{\hbar\ep_2}~,
\ee
with $\text{Re}(\hbar\ep_1)>0$ for $|q|<1$. In practice, the functional property we should use is
\be
\frac{(q^Y;q)_\infty}{(q^X;q)_\infty}\ (1-q)^{Y-X} \equiv \frac{\Gamma_{q}(X)}{\Gamma_{q}(Y)}\xrightarrow{q\to 1}\frac{\Gamma(X)}{\Gamma(Y)}~.
\ee
The 4d Nekrasov function (\ref{eq:4dNek}) can be obtained from the 5d version (\ref{eq:5dNek}) by applying this limit. Similarly, this limiting procedure converts the 3d block integral (\ref{eq:Z3d}) to its 2d version (\ref{eq:Z2d}), without modifying the pole structure and contour prescription.\footnote{\label{foot:DIV}Note that  the limit applied to unnormalized partition functions is generally ill-defined due to overall scaling factors of the type $(1-q)^{...}$. In the limit, we simply ignore such terms.} Also, note that in 5d the instanton charge corresponds to a conserved current, hence the instanton counting parameter can be associated to its fugacity, namely
\be
\mathring{\Lambda}\equiv \e^{-\hbar/g^2_\text{5d}}~,
\ee
where $g^2_\text{5d}$ denotes the dimensionful 5d Yang-Mills coupling. When taking the limit $\hbar \to 0$, it is natural to identify $g^2_\text{5d}\equiv \hbar g^2_\text{4d}$, hence resulting in a week coupling limit.\footnote{Note that in 5d a Chern-Simons level can also be turned on in principle. However, for the purposes of this paper which is mostly focused on 4d, we do not consider such possibility (in the context of BPS/CFT correspondence, this more general case in discussed in \cite{Kimura:2015rgi,Kimura:2019hnw}. Also, in the ADHM matrix model description of 5d instanton partition functions, the Chern-Simons level disappears in the 4d limit.} Correspondingly, the 5d instanton expansion turns into the analogous 4d instanton expansion with fixed instanton counting parameter. A similar reasoning applies between the 3d FI $\zeta$ (the mass for the topological symmetry) and the 2d FI $\xi$. 

As we have reviewed in the first part of this section, instanton/vortex partition functions of the 5d/3d quiver gauge theories have a natural dual description in terms of $\text{W}_{q,t^{-1}}(\Xi)$ correlators. The main goal of this work is to follow the 4d/2d limit we have just described in the dual $\text{W}_{q,t^{-1}}(\Xi)$ algebra side, and we will simply refer to it as the 4d gauge theory limit or $\hbar$-limit for short. We do that in the next section, which contains the main results of the paper.

\section{Quiver $\W_{\ep_1,\ep_2}$ algebras}\label{sec:hWalg}
In this section, we construct an algebra associated to the quiver $\Xi$ which we call $\W_{\ep_1,\ep_2}(\Xi)$. These  type of algebras are naturally associated to 4d $\mathcal{N}=2$ quiver gauge theories described in the previous section. In fact, our approach is parallel to KP construction of the $\text{W}_{q,t^{-1}}(\Xi)$ algebras associated to 5d $\mathcal{N}=1$ theories. For the sake of clarity, we start by giving an axiomatic definition through a free boson construction, explaining how the $\W_{\ep_1,\ep_2}(\Xi)$ can be thought of as the additive version (in spectral parameters) of $\W_{q,t^{-1}}(\Xi)$. Afterwards, we follow a constructive approach, showing  how the former can be obtained through a consistent application of the $\hbar$-limit to the latter. Let us emphasize that this limit is not the usual CFT-limit which recovers, for instance, the standard $\text{W}_N$ algebra from its deformation $\text{W}_{q,t^{-1}}(A_N)$. In fact, the CFT-limit rescales only the deformation parameters ($q,t,\mu_e$) appearing in the deformed Cartan matrix (\ref{eq:qdefC}) while keeping fixed the spectral parameters of the generating and screening currents. For instance, in the case of the $q$-Virasoro algebra discussed around (\ref{eq:qVir}), the CFT-limit applied to the generating current yields
\be
\mb{T}(z)=2+\beta \left(z^2 \mb{L}(z)-\frac{c-1}{24}\right)\hbar^2+\ldots~,
\ee
where $\mb{L}(z)$ is the Virasoro current with central charge  $c=1-6(\sqrt{\beta}-1/\sqrt{\beta})^2$. The effect of this limit on 2-point functions of vertex operators and screening currents is
\be
\frac{(x;q)_\infty}{(q^{\alpha}x;q)_\infty} \xrightarrow{q\to 1}(1-x)^\alpha~,
\ee
leading to the familiar Dotsenko-Fateev representation of conformal blocks in terms of Euler-Selberg type integrals. Here, we are instead interested in rescaling the spectral parameters as well: as recalled at the end of the previous subsection, this is the (weak coupling) limit which preserves the quiver structure of gauge theories upon dimensional reduction. As a result, we obtain a new type of $\ep$-deformed algebra whose Dotsenko-Fateev conformal blocks are expressed in terms of Mellin-Barnes type integrals. In particular cases, we can show that the latter are exactly the same as ordinary CFT conformal blocks upon suitable identifications dictated by spectral duality. Finally, as a further application, we investigate dual Jack functions, which can be characterized as the eigenfunctions of the rational Ruijsenaars-Schneider model dual to the trigonometric Calogero-Sutherland model. These functions arise as the $\hbar$-limit of Macdonald functions and may be associated to some notion of singular vectors of the $\text{W}_{\ep_1,\ep_2}(A_N)$ algebra in the same way as Macdonald/Jack polynomials are for $\text{W}_{q,t^{-1}}(A_N)/\text{W}_N$. 


\subsection{Warming up: $\ep$-Virasoro}

In order to acquire familiarity with the formalism, let us start by
reviewing the definition and the free boson realization of the
simplest $\text{W}_{\ep_1,\ep_2}(\Xi)$ algebra, namely the $\ep$-Virasoro algebra associated to $\Xi=A_1$. It first appeared in \cite{Hou:1996fx}, and we closely follow that presentation. Abstractly, it can  be defined through the quadratic relation satisfied by its current
\begin{multline}\label{eq:ThVirDef}
f(W-Z)\mb{T}(Z)\mb{T}(W)-f(Z-W)\mb{T}(W)\mb{T}(Z)=\\
=2\pi\i\frac{\ep_1 \ep_2}{\epp}\Big(\delta(\epp+W-Z)-\delta(-\epp+W-Z)\Big) ~,
\end{multline}
where we have defined the structure function
\be\label{eq:strFun}
f(X)\equiv \frac{\Gamma_1(-X|2\epp)\Gamma_1(\epp+\ep_2-X|2\epp)\Gamma_1(\epp+\ep_1-X|2\epp)}{\Gamma_1(2\epp-X|2\epp)\Gamma_1(\ep_2-X|2\epp)\Gamma_1(\ep_1-X|2\epp)}~,
\ee
and on the r.h.s. we find the ordinary (additive) Dirac $\delta$ function. We are now able to explain in which sense this algebra can be thought of as the additivization of the $q$-Virasoro algebra: taking its deformation parameters to be ($\beta\equiv-\ep_2/\ep_1$)
\be
q\equiv \e^{-\hbar \ep_1}~,\qquad t\equiv \e^{\hbar \ep_2}~,\qquad p\equiv \e^{-\hbar \epp}~,\qquad \ep_+\equiv \ep_1+\ep_2~,\qquad \mu_e\equiv \e^{-\hbar m_e}~,
\ee
and rescaling all the spectral parameters according to
\be
x\equiv \e^{\hbar X}~,\qquad w\equiv \e^{\hbar W}~,\qquad z\equiv \e^{\hbar Z}~,
\ee
the direct application of the $\hbar\to 0$ limit to  (\ref{eq:qVirdef}) yields the quadratic relation (\ref{eq:ThVirDef}) since $\mathring{f}(x) \to f(X)$ thanks to the property (\ref{eq:gammalimit}). We see that, after the limit, the spectral parameters naturally enter in additive form.

\begin{figure}[!ht]
\leavevmode
\begin{center}
\includegraphics[height=0.20\textheight]{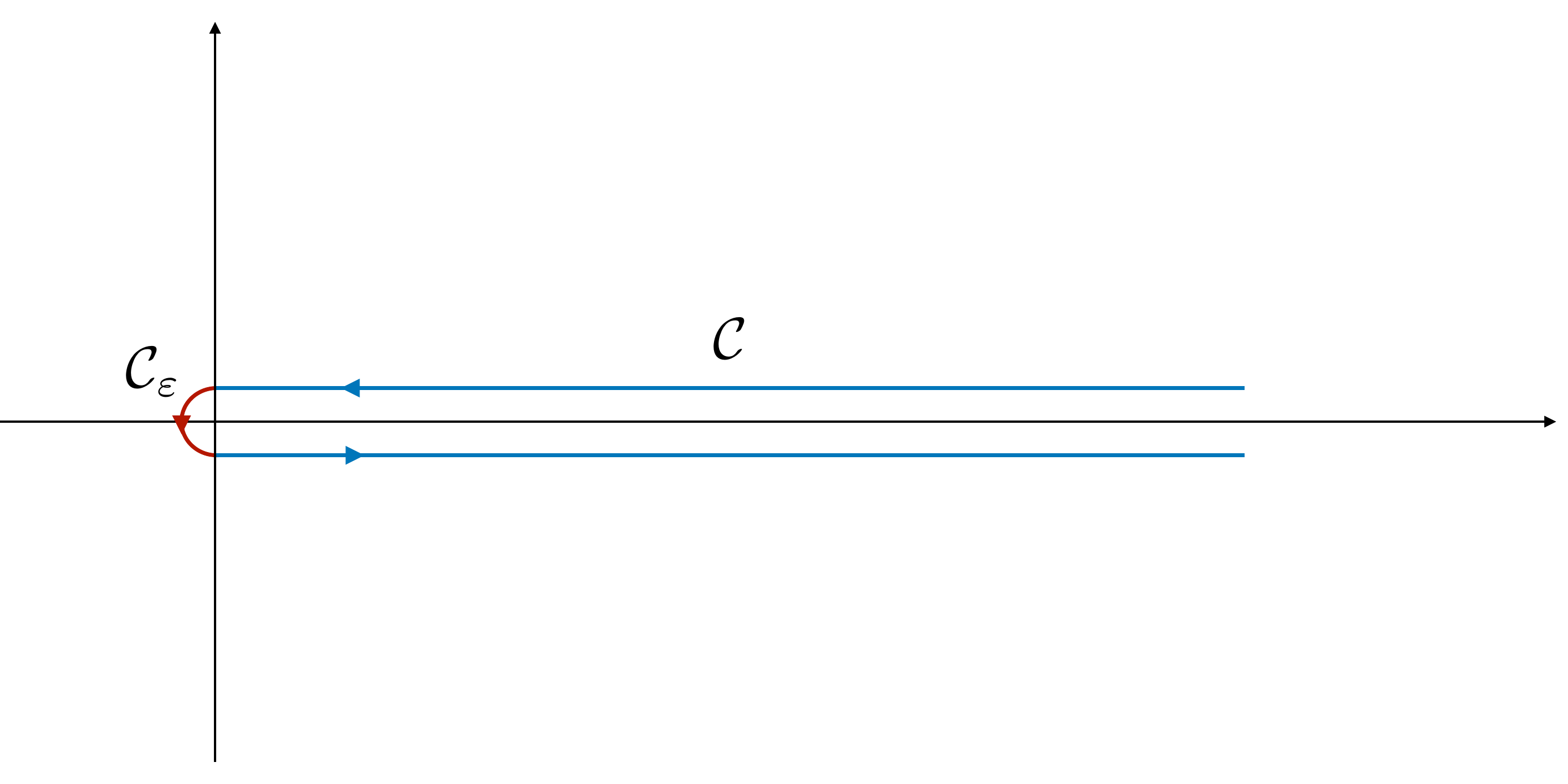}
\end{center}
\caption{The integration contour describing the normal ordering/regularization prescription.}
\label{fig:contour}
\end{figure}

In order to give a concrete realization of the algebra, one needs to specify a representation for the functions involved in the defining relation, in particular we take the following representation of the $\delta$ function
\be
\delta(X)=-\frac{1}{2\pi\i}\left(\frac{1}{X+\i 0}+\frac{1}{-X+\i0}\right)~.
\ee
We assume that the algebra is generated by a current (which we continue to call $qq$-character) 
\be\label{eq:ThVirGen}
\mb{T}(X)\equiv \mb{Y}(\epp/2+X)+\mb{Y}(-\epp/2+X)^{-1}~,
\ee
for which we would like to find a free boson representation. In order to do that, it is instrumental to find the bosonization of the screening current first, which we take to be a free boson operator $\mb{S}(X)$ satisfying 
\be\label{eq:commutantSbis}
 [\mb{T}(Z),\mb{S}(X)]=\frac{\d_{\ep_1}}{\d_{\ep_1} X}\, (\delta(X-Z)\, \cdots)~,
\ee
where we have  introduced the finite $\ep$-derivative
\be
\frac{\d_{\ep}}{\d_{\ep} X}f(X)=\frac{f(X+\ep/2)-f(X-\ep/2)}{\ep}~.
\ee
At this point, it is clear what should be its characterization in terms of the 2-point function: we must have (up to constant proportionality factors)
\be\label{eq:4dSSVir}
\mb{S}(X)\mb{S}(Y)= \ :\mb{S}(X)\mb{S}(Y): \times\frac{\Gamma_1(X-Y-\ep_2|\ep_1)\Gamma_1(X-Y+\ep_1|\ep_1)}{\Gamma_1(X-Y|\ep_1)\Gamma_1(X-Y+\ep_+|\ep_1)}\, \e^{-\gamma\ep_1\ep_2}~,
\ee
which readily follows from the application of the $\hbar$-limit to the 2-point function (\ref{eq:SaSb}) of the $q$-Virasoro algebra. For reasons that will be clarified momentarily, it is very convenient to work in a continuous basis (see \cite{Zenkevich:2017ylb} for an alternative realization):\footnote{\label{foot:NO}This approach seems to directly connect to a double quantization of the profile approach of \cite{Nekrasov:2003rj}.} this fundamental relation (which is all we need for computations) can be bosonized by introducing a Heisenberg algebra generated by oscillators $\mb{s}(k)$ labeled by a continuous real (for the moment) parameter $k\in\mathbb{R}\backslash\{0\}$ and satisfying the commutation relation 
\be
[\mb{s}(k),\mb{s}(k')]=-\frac{4}{k}\sinh(\ep_1 k/2)\sinh(\ep_2 k/2)\delta(k+k')C(k)~,
\ee
where we have defined the deformed and $k$-dependent $A_1$  Cartan matrix
\be
C(k)\equiv 2\cosh(\epp k/2)~.
\ee
The free boson operators we are looking for can now be introduced as
\begin{align}
\mb{S}(X)\equiv &\ : \e^{\mb{s}(X)}:~,\quad \mb{s}(X)\equiv \dashint\d k\, \frac{\mb{s}(k)\, \e^{-k X}}{2\sinh(\ep_1 k/2)}~,\\
 \mb{Y}(X)\equiv &\ :\e^{\mb{y}(X)}:~,\quad \mb{y}(X)\equiv \dashint\d k\, \mb{y}(k)\,\e^{-k X}~,\label{eq:YhVir}
\end{align}
where we set $\mb{s}(k)\equiv C(k)\mb{y}(k)$ and  kept the definition of the normal ordering formally unchanged, namely the integration over the positive domain is pushed to the right of the integration over the negative domain.  However, the dashed integral notation reminds us that we should somehow specify what happens around origin: the choice of the continuous basis requires some regularization to deal with the generally bad behavior of integrals of $\mathbb{C}$-valued functions of the form
\be
\int_0^\infty \frac{\d k}{k} \, \ldots~,
\ee
arising from normal ordering of operator products. Typically, we find a $\ln(\varepsilon)$ divergence for this type of integrals, where $\varepsilon$ denotes the distance from the origin. In order to resolve this problem, we follow the prescription of \cite{Jimbo:1996ev,Jimbo:1996ss} and interpret our integrand as the jump across a cut in the integrand of a well-defined contour integral: we complexify the integration variables, introduce $\ln(-k)$ under the integral with a cut along the positive real half-line and replace the integral over the real half-line with a Hankel contour $\mathcal{C}$ (fig. \ref{fig:contour}), namely 
\be\label{eq:normalprescription}
\dashint_0^\infty\frac{\d k}{k}\ \ldots\to \oint_{\mathcal{C}}\frac{\d k}{2\pi\i k}\ln(-k)\ \ldots = \int_0^\infty\frac{\d k}{k}\ \ldots+\oint_{\mathcal{C}_\varepsilon}\frac{\d k}{2\pi\i k}\ln(-k)\ \ldots ~.
\ee
With this normal ordering/regularization prescription, all the integrals we will have to deal with are well-defined. In fact, the $\ep$-Virsoro algebra is simple enough to explicitly show that the current (\ref{eq:ThVirGen}) in the representation (\ref{eq:YhVir}) satisfies (\ref{eq:ThVirDef}), the key property being the integral representation of the $\Gamma_1$ function recorded in (\ref{eq:Gamma1int}). In the following, we define the $\text{W}_{\ep_1,\ep_2}(\Xi)$ algebras by generalizing this construction and relying on their characterization through the screening currents.

\textbf{Remark}. The regularization prescription which we have adopted seems to be quite \textit{ad-hoc}. However, let us observe that all it does is to remove a divergent constant coming from integrating on a small region around the origin. Similar divergencies arise also in the large $N$ analysis of matrix models \cite{Eynard:2015aea,Livan_2018} and are similarly cured. For the time being, we consider the prescription above as a working hypothesis and postpone further comments to a later subsection.

\subsection{Generalization: arbitrary quivers}
The generalization to an arbitrary quiver $\Xi$ starts by considering the very same deformed Cartan matrix (\ref{eq:4dcartan}) associated to the 4d quiver gauge theory, namely
\be
C(k)_{ab}\equiv 2\cosh(\epp k/2)\delta_{ab}-\sum_{e:a\to b}\exp(\ep_+ k/2- k m_e)-\sum_{e:b\to a}\exp(-\ep_+ k/2+k m_e)~,
\ee
where the operation $C_{ab}\to C(k)_{ab}$ rescales every parameter by $k$. In order to give a defining free boson representation of the $\W_{\ep_1,\ep_2}(\Xi)$ algebra, we  introduce root-type bosons $\{\mb{s}^a(k),\, k\neq 0\}_{a\in\Xi_\circ}$ satisfying the (continuous) Heisenberg algebra
\be\label{eq:contHeis}
[\mb{s}^a(k),\mb{s}^b(k')]=-\frac{4}{k}\sinh(\ep_1 k/2)\sinh(\ep_2 k/2)\delta(k+k')C(k)_{ab}~.
\ee
With these ingredients, we define the screening currents
\be\label{eq:epScreen}
\mb{S}^a(X)\equiv \ : \e^{\mb{s}^a(X)}:~,\quad \mb{s}^a(X)\equiv \dashint\d k\, \frac{\mb{s}^a(k)\, \e^{-k X}}{2\sinh(\ep_1 k/2)}~,
\ee
which can be used to define the generating currents $\mb{T}^a(Z)$ of the algebra as the commutant of the screening currents in the Heisenberg algebra up to total differences, namely
\be\label{eq:commutantS}
\text{W}_{\ep_1,\ep_2}(\Xi)\equiv \{ \mb{T}^a(Z)\}_{a\in\Xi_\circ}\quad \text{such that}\quad [\mb{T}^a(Z),\mb{S}^b(X)]=\delta_{ab}\frac{\d_{\ep_1}}{\d_{\ep_1} X}\, (\delta(X-Z)\, \cdots)~.
\ee

\textbf{Remark}. Due to the manifest $\ep_1\leftrightarrow \ep_2$ symmetry of the algebra, there is actually a second screening current obtained by implementing this transformation on the definitions (\ref{eq:epScreen}),  (\ref{eq:commutantS}). In this paper, we do not consider the dual screening current. 

Alternatively, we may also introduce the weight-type bosons $\{\mb{y}^a(k),\, k\neq 0\}_{a\in\Xi_\circ}$ satisfying the commutation relations
\be
[\mb{y}^a(k),\mb{s}^b(k')]=-\frac{4}{k}\sinh(\ep_1 k/2)\sinh(\ep_2 k/2)\delta(k+k')\delta_{ab}~,
\ee
and define the fundamental vertices
\be
\mb{Y}^a(X)\equiv \ :\e^{\mb{y}^a(X)}:~,\quad \mb{y}^a(X)\equiv \dashint\d k\, \mb{y}^a(k)\,\e^{-k X}~.
\ee
Then, the $\W_{\ep_1,\ep_2}(\Xi)$ currents can also be defined as the $qq$-characters (see \cite{Jeong:2019fgx} for a recent gauge theory application) 
\be\label{eq:hWcurr}
\mb{T}^a(X)\equiv \mb{Y}^a(\ep_+/2+X)+\text{all iWeyl reflections}~.
\ee
For instance, for the single node quiver $\Xi=A_1$ discussed in the previous subsection, one can explicitly verify (appendix \ref{app:screenProp}) that the generating current (\ref{eq:ThVirGen}) fulfills (\ref{eq:commutantS}) or satisfies the quadratic relation (\ref{eq:ThVirDef}). Note that operator products should be understood with a $\i 0$ prescription, namely we assume the operators on the left are slightly above the operators on the right along a chosen direction (real axis).

\subsection{$\hbar$-limit: from $\W_{q,t^{-1}}$ to $\W_{\ep_1,\ep_2}$  algebras}

In this section, we show how the quiver $\W_{\ep_1,\ep_2}(\Xi)$ algebras also arise as a particular limit of the quiver $\W_{q,t^{-1}}(\Xi)$ algebras we have reviewed in section \ref{sec:5dKP}. In fact,  this is exactly the $\hbar$-limit we have already introduced in subsection \ref{subsec:gaugelimit} when discussing the gauge theory side of the story and which has led us to postulate various definitions in previous subsections.

However, we can do better because we can directly implement and follow this limit on the free boson realization of the parent $\text{W}_{q,t^{-1}}(\Xi)$ algebras. Let us start by considering the screening currents (\ref{eq:qdefS}), which we recall here for convenience
\be\label{eq:qdefS}
\S^a(x)= \ :\e^{\s^a(x)}:~,\quad \s^a(x)= -\sum_{n\neq 0}\frac{\s^a_n\ x^{-n}}{q^{n/2}-q^{-
n/2}} +\sqrt{\beta}\s^a_0+\sqrt{\beta}\tilde\s^a_0\ln x~.
\ee
Upon setting $x=\exp(\hbar X)$, the non-zero modes of the field  $\s^a(x)$ may formally be thought of as discrete Laplace modes (or Fourier  depending of the reality conditions) with sampling step $\hbar$. In the limit of the step going to zero, one should switch to a continuous  transform while keeping $\hbar n$ fixed in the limit, namely
\be\label{eq:discTocont}
\hbar n\to k~,\qquad \hbar\sum_{n\neq 0}\to\dashint\d k~,\qquad \frac{1}{\hbar}\mb{s}_n^a\to \mb{s}^a(k)~. 
\ee 
Therefore, for the oscillator part of the field $\mb{s}^a(x)$ one has the $\hbar\to 0$ limit
\be
-\sum_{n\neq 0}\frac{\s^a_n\ x^{-n}}{q^{n/2}-q^{-
n/2}}\  \to \ \dashint \d k\, \frac{\mb{s}^a(k)\, \e^{-k X}}{2\sinh(\ep_1 k/2)}~,
\ee
with the oscillators satisfying the continuous Heisenberg algebra introduced in (\ref{eq:contHeis}), hence reproducing (\ref{eq:epScreen}). Finally, this limiting procedure applied to the $qq$-charactes  (\ref{eq:Tcurr}) yields the $qq$-charactes given in (\ref{eq:hWcurr}). 

\textbf{Remark}. This implementation of the $\hbar$-limit seems to be quite natural at least in the region $|n|\gg1$, where $\hbar n$ can indeed be traded for a continuous label $k$. On the other hand, the treatment around the region $k=0$ requires some care, such as the regularization prescription introduced in (\ref{eq:normalprescription}). In fact, we can now understand why the normal ordering functions would be divergent without any regularization. In taking the $\hbar$-limit of well-defined quantities, we usually strip-off simple factors of $(1-q)^{\ldots}\simeq \exp(\ln(\hbar\ep_1)\ldots)$, which would otherwise lead to undefined limits. In the operatorial picture, the  $\hbar$-limit is implemented by switching to a continuous boson, and then much of the problems are caused by the almost-zero modes around $|k|=0$, for which we need to impose a cutoff at $|k|=\varepsilon\to 0$. The divergence then goes like  $\exp(\ln\varepsilon)$, and we get a match upon identifying $ \varepsilon \sim \hbar$, which is indeed the minimal scale in the identifications (\ref{eq:discTocont}). It is tempting to try to treat more properly such divergent terms within the operatorial formalism too, perhaps considering an additional free boson localized around $k=0$, but we do not pursue this possibility here.

\subsection{Zero modes}

In the implementation of the $\hbar$-limit within the operatorial formalism, we have not discussed what happens to the zero mode operators in (\ref{eq:qdefS}). In fact, there are some sensible choices we can make without spoiling the commutation relations (\ref{eq:commutantS}). If we want to preserve the commutator $[\tilde\s^a_0,\s^b_0]=\mathring{C}^{[0]}_{ab}=C(0)_{ab}$, we can either scale 
\be
\frac{1}{\hbar}\s^a_0\to \s^a_0~,\quad \hbar\tilde\s^a_0\to \tilde\s^a_0~,
\ee
or viceversa. The first choice seems more natural and adds the following modes to the various fields
\be
\s^a(X)\to \s^a(X)+\sqrt{\beta}\tilde\s^a_0X~,\quad \y^a(X)\to \y^a(X)-\i\sqrt{\ep_1\ep_2}\tilde\y^a_0~.
\ee
For the second choice, only $\sqrt{\beta}\s^a_0/\hbar$ would appear in the screening currents, hence resulting in a divergent result (which may be reabsorbed into the divergencies we have already mentioned). Actually, there is a third option consisting in not rescaling the zero modes at all: in this case, only $\sqrt{\beta}\s^a_0$ would appear in the screening currents. Moreover, we can always multiply the screening currents by an operator-valued periodic function of period $\ep_1$, which does not spoil the defining property (\ref{eq:commutantS}) provided that the function commutes with the $\text{W}_{\ep_1,\ep_2}(\Xi)$ currents. The minimal choice can be made by going back to (\ref{eq:qdefS}) and replacing
\be
\sqrt{\beta}\tilde\s^a_0\ln x\to \ln \frac{\Theta(q^\lambda x q^{-\sqrt{\beta}\tilde\s^a_0};q)}{\Theta(q^\lambda x ;q)}~,
\ee
for some auxiliary $\lambda$. This can be done because it amounts to multiplying the screening currents by a $q$-constant depending on $\tilde\s^a_0$ only, which commutes with $\T^a(x)$. Then, without rescaling the zero modes, in the $\hbar$-limit we would get
\be
\sqrt{\beta}\s^a_0+\ln \frac{\Theta(q^\lambda x q^{-\sqrt{\beta}\tilde\s^a_0};q)}{\Theta(q^\lambda x ;q)}\to \sqrt{\beta}\s^a_0+\ln \frac{\sin\frac{\pi}{\ep_1}(\ep_1\lambda-X -\ep_1\sqrt{\beta}\tilde\s^a_0)}{\sin\frac{\pi}{\ep_1}(\ep_1\lambda -X)}~.
\ee
A suitable combination of the choices above may also work, but for our purposes we  consider only the first, simplest possibility.

\subsection{Correlation functions: instantons and vortices}

We can now move to compute the main objects of our interests, namely correlators of free boson vertex operators and integrated screening currents. In order to begin with, let us compute the basic building block, namely the 2-point function of the screening currents
\begin{align}\label{eq:4dSS}
\mb{S}^a(X_i)\mb{S}^b(X_j)=& \ :\mb{S}^a(X_i)\mb{S}^b(X_j): \times \exp\Big[\oint_\mathcal{C}\frac{\d k}{2\pi\i k}\ln(-k)\frac{\e^{-(X_i-X_j)k}\sinh(\ep_2 k/2)}{\sinh(\ep_1 k/2)}\, C(k)_{ab} \Big]=\nn\\
=&\ :\mb{S}^a(X_i)\mb{S}^b(X_j):\times\left[\frac{\Gamma_1(X_{ij}-\ep_2|\ep_1)\Gamma_1(X_{ij}+\ep_1|\ep_1)}{\Gamma_1(X_{ij}|\ep_1)\Gamma_1(X_{ij}+\ep_+|\ep_1)}\, \e^{2\gamma\beta}\right]^{\delta_{ab}}\nn\times\\
&\ \times\prod_{e:a\to b}\frac{\Gamma_1(X_{ij}+m_e|\ep_1)}{\Gamma_1(X_{ij}+m_e-\ep_2|\ep_1)}\, 
\e^{-\gamma\beta} \prod_{e:b\to a}\frac{\Gamma_1(X_{ij}-m_{e}+\ep_+|\ep_1)}
{\Gamma_1(X_{ij}-m_e+\ep_1|\ep_1)}\, \e^{-
\gamma\beta}~,
\end{align}
where we have set $X_{ij}\equiv X_i-X_j$ and used the integral representation (\ref{eq:Gamma1int}).  We can now generalize to the entire quiver with an arbitrary number 
$r_a$ of insertions for each type, namely
\begin{multline}
\prod_{a=1}^{|\Xi_\circ|}\prod_{i=1}^{r_a} \S^a(X^a_{i})=\ :\prod_{a=1}^{|\Xi_\circ|}\prod_{i=1}^{r_a} \S^a(X^a_{i}): \prod_a c^a_{\ep_1}(X^a)\times\prod_{a}\Delta^a_\text{nodes}(X^a)\prod_{a<b}\Delta^{ab}_\text{arrows}(X^{a}-X^{b})~,
\end{multline}
where we have defined the functions
\begin{align}\label{eq:hDeltas}
\Delta^a_{\text{nodes}}(X^a)\equiv &\ \prod_{\substack{1\leq i\neq j\leq r_a}} \frac{\Gamma_1(-\ep_2+X^a_{ij}|\ep_1)}{\Gamma_1(X^a_{ij}|\ep_1)}\prod_{e:a\to a}\frac{\Gamma_1(m_e+X^a_{ij}|\ep_1)}{\Gamma_1(-\ep_2+m_e+X^a_{ij}|\ep_1)}  ~,\nn\\
\Delta^{ab}_{\text{arrows}}(X^a-X^b)\equiv&\  \e^{\beta\gamma C(0)_{ab}r_a r_b}\prod_{\substack{i = 1,...,r_a\\ j=1,\ldots, r_b}}\prod_{e:a\to b}\frac{\Gamma_1(X^a_i-X^b_j+m_e|\ep_1)}{\Gamma_1(X^a_i-X^b_j+m_e-\ep_2|\ep_1)}\times\nn\\
& \ \times \prod_{e:b\to a}\frac{\Gamma_1(X^a_i-X^b_j-m_{e}+\ep_+|\ep_1)}
{\Gamma_1(X^a_i-X^b_j-m_e+\ep_1|\ep_1)}~,\nn\\
c^a_{\ep_1}(X^a)\equiv&\ \e^{\beta\gamma C(0)_{aa}\frac{r_a(r_a-1)}{2}} \prod_{\substack{i<j}}\frac{S_1(-\ep_2+ X^a_{ji}|\ep_1)}{S_1(X^a_{ji}|\ep_1)}\prod_{e:a\to a}\frac{S_1(m_e +X^a_{ji}|\ep_1)}{S_1(-\ep_2+ m_e+X^a_{ji}|\ep_1)}~.
\end{align}
Note that the last function is the $\hbar$-limit of a $q$-costant, namely a periodic function under shifts $X\to X\pm \ep_1$, and therefore we can neglect it for most of our purposes.\footnote{We recall that, because of the screening property (\ref{eq:commutantS}), the insertion of $\ep_1$-periodic functions cannot be uniquely fixed by algebraic considerations alone.}

Finally, the operators above are taken to act on the Fock space built on a (charged) Fock vacuum $\ket{\alpha}$ and its dual $\bra{\alpha}$, which are characterized by
\begin{align}
\s^a(k>0)\ket{\alpha}&=0~,\quad \tilde\s_0^a\ket{\alpha}=\alpha^a\ket{\alpha}~,\quad \ket{\alpha}
\equiv\e^{\sum_{a}\alpha^a \y_0^a}\ket{0}~,\nn\\
\bra{\alpha}\s^a(-k)&=0~,\quad \bra{\alpha}\tilde \s_0^a=\alpha^a\bra{\alpha}~,\quad \bra{\alpha}
\equiv\bra{\alpha}\e^{-\sum_{a}\alpha^a \y_0^a}~,
\end{align}
where we assume $\braket{0}{0}=1$. As in the $q$-deformed case, the Fock charges will parametrize the instanton/vortex counting parameters of the gauge theories. Note that this is radically different w.r.t. the AGT perspective, where coupling constants are identified with insertion points of vertex operators.

In the following, we are going to show how to use these basic building blocks to construct 4d instanton and 2d vortex partition functions of gauge theories which in subsection \ref{sec:4d2dgauge} we have associated to the very same quiver $\Xi$. In fact, the construction is going to be essentially unchanged w.r.t. KP construction which we have reviewed in subsections \ref{subsec:Qinfinity}, \ref{subsec:5dGT}, with the major difference that we should switch to an additive notation adapted to the new operators. Therefore, we will be quite brief.

\subsubsection*{4d $\mathcal{N}=2$ theories: infinitely-many screening charges.}

Following our discussion in subsection \ref{subsec:Qinfinity}, we are mainly interested in defining the $\mb{Z}$ state for the $\W_{\ep_1,\ep_2}(\Xi)$  algebra. In view of the screening relation (\ref{eq:commutantS}), we consider screening charges defined through an additive version of Jackson integrals, namely 
\be\label{def:addjackS}
\Q^a_Z\equiv \int_Z\d X_{\ep_1}\ \S^a(X)\equiv \sum_{k\in\mathbb{Z}}  \S^a(Z-k\ep_1)~.
\ee
The set of points $\chi_\emptyset \equiv \cup_{a\in\Xi_\circ}\chi^a_{\emptyset}$ describing the ground configuration is chosen to be
\be
\chi^a_{\emptyset}\equiv \{Z_{\emptyset^a_{Ai}}\equiv X_{aA}+\ep_2(1-i) | A = 1, ..., N_a \ , i = 1,..., \infty\}~,
\ee
where $a$ runs over the nodes of the quiver and $N_a$ denotes the rank of the node. We similarly define the set $\chi \equiv \cup_{a\in\Xi_\circ}\chi^a$ in which each base point is arbitrarily shifted by $\ep_1$ according to a sequence of integers $k^a_{Ai}\in\mathbb{Z}$ (excited configurations), namely
\be
\chi^a\equiv \{Z_{k^a_{Ai}}\equiv Z_{\emptyset^a_{Ai}} - k^a_{Ai}\ep_1\ | A = 1, ..., N_a \ , i = 1,..., \infty\}~ .
\ee
We next define the operator  
\be\label{eq:Zoperator}
\widehat{\mb{Z}}\equiv  \prod^{\prec}_{Z\in \chi_\emptyset}\Q^{\i(Z)}_Z=\sum_{\{k^a\}} \prod_{a=1}^{|\Xi_\circ|}\prod^{\prec}_{Z\in \chi^a}  \S^{a}(Z)  ~,
\ee
where  the ordering is inherited by the one in exponentiated variables. As we know, the sum over integer sequences can actually be recast as a sum over partitions, and one of our main results is that the state
\be
\ket{\widehat{\mb{Z}}}\equiv \widehat{\mb{Z}}\ket{\widehat\alpha_0}
\ee
corresponds to the (unnormalized) time-extended instanton partition function of the quiver gauge theory. For instance, by applying the representation (\ref{eq:4dNek}) and the 2-point function (\ref{eq:4dSS}) to the simplest $\Xi=A_M$ linear quiver, we have that the projected state $\braket{\widehat\alpha_\infty}{\widehat{\mb{Z}}}$ matches (\ref{eq:4dAM}) once normalized, with the instanton counting parameters identified with the Fock momenta $\Lambda_a\equiv \exp(-\sqrt{\beta}\ep_1\widehat\alpha^a_0)$ and Coulomb branch parameters identified with the variables $X_{aA}$ parametrizing the strings of base points. Coupling to (anti-)fundamental matter, associated to the presence of additional arrows with either no target or source, can be included by giving a background to the negative oscillators/time variables of the form 
\be
-\frac{\s^a(-k)}{2\sinh(\ep_1 k/2)}\simeq \tau^a(k) \to \tau^a(k)-\sum_f\frac{\e^{-k m^a_f}}{4k\sinh(\ep_1 k/2)\sinh(\ep_2 k/2)}~,\quad k>0~,
\ee
with $m_f$ encoding the fundamental mass parameters, or (more generally) by the insertion of additional vertex operators
\be
\mb{V}^a(m)\equiv \ :\e^{\mb{v}^a(m)}:~,\quad \mb{v}^a(\mu)\equiv \sum_{n\neq 0}\frac{\mb{y}^a(k)\, \e^{-m^a k}}{4\sinh(\ep_1 k/2)\sinh(\ep_2 k/2)}~,
\ee
whose effect is inserting the appropriate $n_{\emptyset k^a_A}(-X_{aA}+m)$ or $n_{k^a_A\emptyset}(X_{aA}-\bar m)$ factors in the instanton partition function. It would be interesting to relate the Ward identities resulting from our construction to the recursion relations of \cite{Kanno:2013aha}.

\textbf{Remark}. As we have already recalled when discussing the 5d setup, from the perspective of the 4d $\mathcal{N}=2$ quiver gauge theory the time-extension of the partition function refers to the possibility of perturbing the prepotential $\mathcal{F}$ in the UV by holomorphic gauge invariant polynomials (higher Casimirs) of the vector multiplet scalar $\Phi$, that is
\be
\mathcal{F}\to \mathcal{F}+\sum_a\sum_{j> 0}\tau^a_j \,\textrm{tr}\, \Phi_a^j~.
\ee
In our formalism, these terms must be encoded by the surviving negative oscillators in the $\mb{Z}$ state acting on the Fock vacuum, namely
\be
-\int_{>0} \d k\, \frac{\mb{s}^a(-k)\, \e^{k X}}{2\sinh(\ep_1 k/2)}\simeq \int_{>0} \d k\, \tau^a(k)\, \e^{k X}~.
\ee
Besides being continuous, this exponential parametrization is more natural for the 5d theory on a circle rather than for its 4d limit (see however footnote \ref{foot:NO}). We are thus led expand the exponential and define discrete time variables through a Mellin-like transform 
\be
\tau^a_j\equiv \frac{1}{j!}\int_{>0}\d k\, k^j\, \tau^a(k)~.
\ee 
In fact, one can try to introduce a discrete basis directly at the level of the Heisenberg algebra, for instance $\mb{s}^a(X)\simeq \sum_{j\geq 0}(\mb{a}^{a}_{j+}+(-1)^j\mb{a}^{a}_{j-}) X^j/j!$ with
\be\label{eq:sMellin}
\mb{a}^{a}_{j+}\simeq (-1)^j\int_{>0}\d k\, k^j\, \frac{\s^a(k)}{2\sinh(\ep_1 k/2)}~,\quad \mb{a}^{a}_{j-}\simeq -\int_{>0}\d k\, k^j\, \frac{\s^a(-k)}{2\sinh(\ep_1 k/2)}~.
\ee
We elaborate more on this possible correspondence in appendix \ref{sec:cont-discrete}. This type of transform may be the bridge to establish a direct connection to earlier works investigating quantum (affine Yangian) algebras and 4d $\mathcal{N}=2$ theories \cite{maulik2012quantum,schiffmann2012cherednik,alex2014affine,Bourgine:2018uod}, but we do not develop this topic here. However, the commutation relations in this basis are more involved, hence this representation is less appealing as far as actual computations are concerned: this is the price one would have to pay to work with a more standard algebraic setup based on discrete oscillators and rational rather than trigonometric series. This tension (simple vs. standard) is at the origin of the more sophisticated tools needed to approach 4d theories through $\W_{\ep_1,\ep_2}$ algebras compared to the 5d or $q$-deformed counterparts.

\subsubsection*{2d $\mathcal{N}=(2,2)$ theories: finitely-many screening charges.}

In this case it is also useful to explicitly introduce the type of vertex operators we are interested in, namely
\be
{\mb H}^a(m|\bar m)\equiv \ :{\mb V}^a(m-\ep_1/2){\mb V}^a(\bar m-\ep_1/2)^{-1}:~,
\ee
whose 2-point function with the screening current reads
\be
{\mb H}^a(Y|\bar Y){\mb S}^b(X)=\ :{\mb H}^a(Y|\bar Y){\mb S}^b(X):\, \left[\frac{\Gamma_1(Y-X|\ep_1)}{\Gamma_1(\bar Y-X|\ep_1)}\, \right]^{\delta_{ab}}~.
\ee
With these ingredients, we can give a dual algebraic interpretation of the disk partition function of the quiver gauge theories, the main object being the state
\be
\mb{Z}\equiv \prod_{a}\left(\oint \d X \mb{S}^a(X)\right)^{r_a}~,
\ee
involving only a finite number $r_a$ of screening currents for each node of the quiver. As the simplest example, let us consider the single node quiver $\Xi=A_1$ associated to  the $\ep$-Virasoro algebra, for which we take the correlation function involving $N$ vertex operators and a single integrated screening current
\be
\bra{\alpha_\infty}\prod_{j=1}^N{\mb H}(Y_j|\bar Y_j)\oint\d X\, {\mb S}(X)\ket{\alpha_0}\simeq \oint\d X\, \e^{\sqrt{\beta}\alpha_0 X}\,\prod_{j=1}^N\frac{\Gamma_1(Y_j-X|\ep_1)}{\Gamma_1(\bar Y_j-X|\ep_1)}~,
\ee
where we neglected a  proportionality factor. This expression can immediately be matched with the Coulomb branch expression of the disk partition function of the 2d $\mathcal{N}=(2,2)$ SQED given in subsection \ref{sec:4d2dgauge}. If we consider the contour to take the contribution from the poles at $X=Y_i+n\ep_1$, $n\in\mathbb{Z}_{\geq 0}$, this correlation function can be written in terms of the hypergeometric series, namely\footnote{We denote ${}_2F_1[a,b;\ep,c|z]={}_2F_1[a/\ep,b/\ep;c/\ep|z]$. Also, if $f(x)$ has a pole at $x=x_0$, $f(x_0)$ is its residue.}
\begin{multline}
\bra{\alpha_\infty}\prod_{j=1}^N{\mb H}(Y_j|\bar Y_j)\oint_{\mathcal{C}_i}\d X\, {\mb S}(X)\ket{\alpha_0}\simeq\\
\simeq z^{Y_i/\ep_1}\, \prod_{j=1}^N\frac{\Gamma_1(Y_j-Y_i|\ep_1)}{\Gamma_1(\bar Y_j-Y_i|\ep_1)}\, {}_N F_{N-1}{\tiny\left[\begin{array}{cc}Y_i-\bar Y_i+\ep_1,&Y_i-\bar Y_j+\ep_1 \\ \ep_1,&Y_i-Y_j+\ep_1\end{array}|z\right]}~,
\end{multline}
where $z\equiv \exp(\sqrt{\beta}\alpha_0 \ep_1)$. In particular, the hypergeometric series capture the non-perturbative vortex contribution to the partition function.

\section{$\text{W}_N/\text{W}_{\ep_1,\ep_2}(A_{M-1})$ spectral duality}\label{sec:spectralD}

In this section, we provide some more explicit comparisons with various types of conformal blocks (in free field formalism or not) which can be computed in 2d CFT with ordinary $\text{W}_N$ symmetry. Since our algebras are supposed to compute the same objects, this is a test of the proposed spectral duality between the two descriptions. We initiate with degenerate Toda conformal blocks featuring in the AGT correspondence\footnote{The strong form of the correspondence involves Liouville/Toda correlators and compact space partition functions \cite{Pestun:2007rz,Hama:2012bg,Benini:2012ui,Doroud:2012xw}. In order to adapt our setup to such observables, we should implement the modular double construction along the lines of \cite{Nedelin:2016gwu}.} and then analyze spectral duality of known integrable systems and relate it to our constructions. 

\subsection{Toda conformal blocks}

The examples we have discussed in the previous subsection allow us to make an interesting comparison with ordinary $\text{W}_N$ correlators involving degenerate insertions and the AGT relation. This restriction is required for having a Lagrangian description of the dual gauge theories, which is not generically the case for class S-theories. For instance, according to the AGT correspondence, the 4-point sphere conformal blocks
\be
\langle V_{\alpha_4}(\infty)V_{\alpha_3=\lambda h_{N-1}}(1)V_{\alpha_2=\kappa h_1}(z)V_{\alpha_1}(0)\rangle~
\ee
in $A_{N-1}$ Toda with two generic primaries (full punctures) and two semi-degenerates\footnote{The $h_i$ denote here fundamental $A_{N-1}$ weights, while $\alpha_{4,1}$ are $N-1$ vectors of generic momenta.} (simple punctures) is captured by the 4d $\mathcal{N}=2$ $\text{SU}(N)$ SQCD partition function \cite{Wyllard:2009hg}, which we have also constructed in terms of $\ep$-Virasoro free boson correlators with $N+2$ insertions. Moreover, a 4d gauge theory can be enriched with BPS surface operators, the simplest case being  a coupled 4d/2d system whose non-trivial dynamics can be described by the 2d $\mathcal{N}=(2,2)$ SQED on the defect (coupled to a free bulk). In the Toda description, this occurs when one of the  semi-degenerate is tuned to be fully degenerate \cite{Alday:2009fs,Kozcaz:2010af,Bonelli:2011wx,Doroud:2012xw}. It is known that in presence of fully-degenerate insertions, the conformal blocks obey additional constraints/decoupling equations \cite{Fateev:2007ab,Fateev:2008bm} which allow them to be expressed in terms of hypergeometric functions (up to conformal factors), namely\footnote{In our notation, $b^2=-\beta$, and $Q_N\equiv (b+b^{-1})\rho$, where $\rho$ is the Weyl vector.}
\begin{multline}
\langle V_{\alpha_4}(\infty)V_{\alpha_3=\lambda h_{N-1}}(1)V_{\alpha_2=-b h_1}(z)V_{\alpha_1}(0)\rangle = z^{\ldots}(1-z)^{\ldots}\times\\
\times  z^{b(\alpha_1-Q_N,h_s)} \ \HFN{(\alpha_4-Q_N,h_k)+\frac{\lambda}{N}-b\frac{N-1}{N}+(\alpha_1-Q_N,h_s)|_{1\leq k \leq N}}{b^{-1},~b^{-1}+(\alpha_1-Q_N,h_s)-(\alpha_1-Q_N,h_k)|_{1\leq k\neq s\leq N}}{z}~,
\end{multline}
for each choice of $s$-channel internal momentum labeled by $1\leq s\leq N$. This coincides with the SQED vortex partition function through the standard AGT dictionary, a result that we have also reproduced from the dual $\ep$-Virasoro perspective involving $N+2$ free boson vertex operators. The map between parameters on dual sides is straightforward, perhaps the most interesting phenomenon to observe once again is the exchange between the cross-ratio and external momentum in the interpretation of the instanton/vortex counting parameter. 

\begin{figure}[!ht]
\leavevmode
\begin{center}
\includegraphics[height=0.8\textwidth]{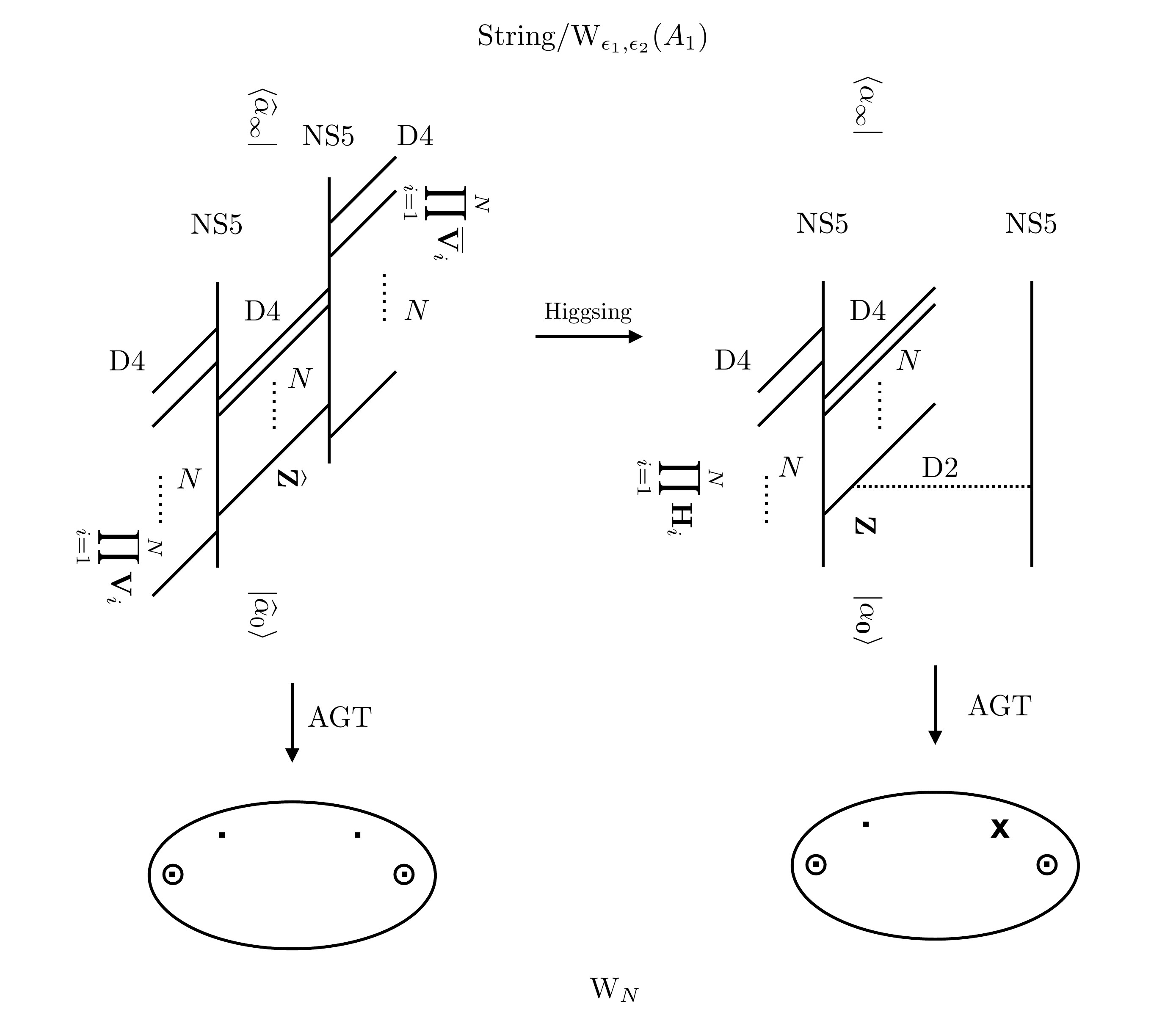}
\end{center}
\caption{Brane picture of the Gauge/$\text{W}_{\ep_1,\ep2}$ correspondence for the simplest quiver $\Xi=A_1$. The lower part refers to the AGT/class-S perspective.}
\label{fig:stringWEB}
\end{figure}

The picture above finds a nice and simple description in terms of brane engineering in string theory (Fig. \ref{fig:stringWEB}). In order to construct 4d $\mathcal{N}=2$ $\text{SU}(N)$ SQCD, our reference example, we can start from a type IIA Hanany-Witten setup involving $M=2$ NS5 intersecting with $N$ D4, with overlapping world-volumes along $\mathbb{C}^2$ \cite{Witten:1997sc}. On the one hand, the dynamical $\text{SU}(N)$ gauge symmetry is supported by the compact parallel segments of the $N$ D4 suspended between the NS5, with their relative positions parametrizing the Coulomb branch parameters and the distance between the NS5 the gauge coupling. On the other hand, in our algebraic construction we can think of them as supporting an infinite number of screening charges of $\text{W}_{\ep_1,\ep_2}(A_{M-1})|_{M=2}=\ep\text{-Virasoro}$. Similarly, the $2N$ semi-infinite D4 give raise to fundamental/anti-fundamental flavors, which in the algebraic side are associated to $N$ vertex operators whose positions and momenta are encoded into their relative positions w.r.t. the compact D4. This brane setup can be Higgsed by aligning for example the rightmost non-compact D4 to the internal ones and pulling the NS5 out of the original plane while stretching a single D2. This will support the $\text{U}(1)$ gauge symmetry of the resulting 2d $\mathcal{N}=(2,2)$ SQED, while in the algebraic description it can be associated to the insertion of a single $\ep\text{-Virasoro}$ screening charge. 
The brane picture is also useful for understanding in which sense  the AGT relation happen to be orthogonal (literally) to our description: the stack of $N$ D4 support the higher rank algebra $\text{W}_N$, with each NS5 intersecting the D4 describing a rank $N-1$ full puncture and a rank $1$ simple puncture. After Higgsing, the emerging D2 is associate to the simple puncture which has turned to fully-degenerate. 

Generally, for 4d $\mathcal{N}=2$ $A$-shaped quivers and unitary gauge groups, the duality we have just described can be generalized to an equivalence of $(M+2)$-point functions in $\text{W}_N$ and $(N+2)$-point functions in $\text{W}_{\ep_1,\ep_2}(A_{M-1})$. We also expect that a similar relation will hold for a wider class of surface defects as well \cite{Gomis:2016ljm}. In the 5d/$q$-deformed setup, the origin of such interplay is known from the string perspective due to fiber/base/S-duality \cite{Katz:1997eq,Bao:2011rc}, the algebraic viewpoint because of Miki's automorphism of the DIM algebra, and the theory of integrable systems: the effect of the duality on the gauge theory side is simply to exchange the rank of the quiver with the rank of the gauge group, with a Lagrangian description existing in both frames; at the quantum group level, the duality simply exchanges two algebras in the same family and their correlation functions; in integrable systems of fully trigonometric type, spectral duality acts within the same class. Instead, in the 4d/$\ep$-deformed setup the duality changes the type of algebras involved, and the very existence of the duality is far less understood. This is because the existence of only one frame where a conventional gauge theory description seems to be valid. In fact, at codim-2, the duality has been shown \cite{Zenkevich:2017ylb} to look very much as Hori-Vafa mirror symmetry between gauged linear sigma models and Landau-Ginzburg theories \cite{Hori:2000kt,Aganagic:2001uw,Gomis:2012wy}. On the algebraic side, the relation between our approach and AGT is ultimately to be found in the representation theory of the affine Yangian which underlies both constructions.

\subsection{rRS model and dual Jack functions}\label{sec:rational-rs-model}

In this last subsection, we would like to explore possible relations between $\text{W}_{\ep_1,\ep_2}(\Xi)$ algebras associated to 4d/2d quiver gauge theories and quantum integrable systems. For simplicity, we focus on $A$-type quivers only. A natural starting point in this direction is the observation that the $\text{W}_{q,t^{-1}}(A_N)$ algebras were introduced in connection with the theory of Macdonald polynomials: in the CFT-limit $q\to 1$, these reduce to Jack polynomials which are in one-to-one correspondence with $\text{W}_N$ singular vectors, while the former turns out to be in one-to-one correspondence with $\text{W}_{q,t^{-1}}(A_N)$ singular vectors \cite{Awata:1995np,Awata:1995zk}. Moreover, another characterization of these orthogonal polynomials is through the integrable systems they are related to: Macdonald or Jack polynomials are eigenfunctions of the trigonometric Ruijsenaars-Schneider (tRS) or Calogero-Moser-Sutherland (tCMS) models respectively.\footnote{More precisely, they describe excitations over the vacuum. The vacuum wavefunction is simply a trigonometric $q/\beta$-deformed Vandermonde determinant which can be stripped off.} The connection to $\text{W}_{q,t^{-1}}(A_N)$ and $\text{W}_N$ algebras goes through the bosonization formulas of the quantum Hamiltonians: once the Heisenberg oscillators are represented on the Fock space of symmetric functions by choosing the power sums basis, the non-diagonal part of the Hamiltonians can be expressed in terms of the algebra generators which annihilate the singular vectors constructed from the screening currents. Here, we aim at applying part of this logic to the case of the $\text{W}_{\ep_1,\ep_2}(A_N)$ algebra.

We are going to show that: $i)$ certain $\text{W}_{\ep_1,\ep_2}(A_N)$ free boson correlators are eigenfunctions of the rational Ruijsenaars-Scheider (rRS) system; $ii$) spectral duality between $\text{W}_{\ep_1,\ep_2}(A_N)$ and $\text{W}_N$  correlators corresponds to the PQ-duality exchanging coordinates and momenta between the rRS and tCMS models.\footnote{Actually, the more appropriate model is the hyperbolic one. From our perspective, the difference amounts in the interpretation of the oscillators either as trigonometric/hyperbolic variables. The reality conditions become important when discussing the actual spectrum of the Hamiltonians, but they play a little role in the bosonization formulas as we always work in a complexified setup.} 
The first relation follows from what we have just recalled, namely  that certain screening integrals of the
$\text{W}_{q,t^{-1}}(A_N)$ algebra are tRS eigenfunctions. Upon taking the $\hbar$-limit, one gets the singular vectors\footnote{By analogy with the $q$-deformed case, we continue to use the world singular vector even though we do not know the exact relation with the representation theory of the $\text{W}_{\ep_1,\ep_2}(A_N)$ algebra.} of the $\text{W}_{\ep_1,\ep_2}(A_N)$ algebra (dual Jack functions) on the one side, and the
corresponding rRS eigenfunctions on the other side. The second relation describes how the spectral/PQ-duality manifests itself on
the level of integrable systems.\footnote{In the context of 3d quiver gauge theories and tRS model, this and other aspects are extensively discussed in \cite{Gaiotto:2013bwa,Bullimore:2014awa,Koroteev:2019byp}.} The tRS model is known
to be PQ-self-dual as the dependance on coordinates and
momenta is trigonometric/trigonometric. On the one hand, after the $\hbar$-limit the tRS Hamiltonian turns into the rRS Hamiltonian in
which the dependance on coordinates becomes rational (the circle on which the coordinates live is decompactified). On the other hand, the tCMS model arises in the dual CFT-limit of tRS, in which it is the dependance on momenta becoming 
rational. Given the action of PQ-duality, it is natural that the two different limits of
tRS are dual to each other: this is the remnant of the self-duality which is broken by the limiting procedures.

In order to develop these points, let us start by recalling the Macdonald/tRS Hamiltonian for a system of $N$ particles with positions $x_i$, which we take to be
\begin{equation}
H_\text{t}\equiv\sum_{i=1}^N\prod_{j\neq i}\frac{tx_i-x_j}{x_i-x_j}\ q^{x_i\partial_{x_i}}~,
\end{equation}
which in the $\hbar$-limit reduces to the rRS Hamiltonian
\begin{equation}
H_2\equiv \sum_{i=1}^N\prod_{j\neq i}\frac{\ep_2+X_i-X_j}{X_i-X_j}\ \e^{-\ep_1\partial_{X_i}}.
\end{equation}
It is an interesting problem to express this operator in terms of $\text{W}_{\ep_1,\ep_2}(A_N)$ generators, as our construction suggests, but we do not attempt to do that here. Instead, our goal is to algorithmically construct its eigenfunctions in terms of the $\text{W}_{\ep_1,\ep_2}(A_N)$ screening charges and compute the resulting eigenvalues. Let us begin with a trivial example, namely the 1-body Hamiltonian whose eigenfunction is
\begin{equation}
\psi(X_1|z_1)\sim z_1^{-X_1/\ep_1}~,
\end{equation}
with eigenvalue $E_2=z_1$. This result holds up to multiplication by a periodic function of the coordinate with period $\ep_1$. As a less trivial yet simple example, let us take the 2-body Hamiltonian, for which it  can be verified that\footnote{We recall that, if a function $f(x)$ has a pole at $x=x_0$, then $f(x_0)$ denotes its residue.}
\begin{multline}
  \psi(X_1,X_2|z_1,z_2) \sim z_2^{-\frac{X_1 + X_2}{\epsilon_1}} \left( \frac{z_1}{z_2} \right)^{-\frac{X_1}{\ep_1}}\
  \frac{\Gamma_1(0|\ep_1)\Gamma_1(X_2-X_1|\ep_1)}{\Gamma_1(-\ep_2|\ep_1)\Gamma_1(-\ep_2+X_2-X_1|\ep_1)}\times\\
\times  {}_2F_1\left[\begin{smallmatrix} \epsilon_+ ,&\epsilon_++X_1-X_2\\
        \ep_1,&\ep_1+X_1-X_2\end{smallmatrix} \left| \frac{z_2}{z_1} \right. \right]
\end{multline}
is an eigenfunction with eigenvalue $E_2=z_1+z_2$. This directly follows from the hypergeometric equation and contiguous recurrence relations. Due to the symmetry of the problem, we can of course get another solution by exchanging $X_1\leftrightarrow X_2$, and we can give an integral representation of the solutions as follows
\begin{equation}
\psi(X_1,X_2|z_1, z_2)\equiv z_2^{-\frac{X_1 + X_2}{\epsilon_1}}\oint_{\mathcal{C}_i} \frac{\d s}{2\pi\i}\ \left( \frac{z_1}{z_2} \right)^{
  -\frac{s}{\ep_1}}\prod_{j=1}^2\frac{\Gamma_1(X_j-s|\ep_1)}{\Gamma_1(-\ep_2+X_j-s|\ep_1)}~, \label{eq:23}
\end{equation}
where the enclosed poles are at $s=X_i+n\ep_1$, $n\in\mathbb{Z}_{\geq 0}$. These two expressions are nothing but a basis of the
4-point functions with two degenerate fields of the
$\ep$-Virasoro algebra, constructed using a single screening current. We have already shown that such free field conformal
blocks are spectral dual to the conformal blocks of the ordinary Virasoro algebra, which are in turn eigenfunctions of the 2-body tCMS system as we recall below.

More generally, the $N$-body eigenfunctions can be obtained by induction using the kernel function for the rRS Hamiltonian\footnote{In the trigonometric case, similar kernel functions arise from completeness of Macdonald/Jack/Schur functions. It is an interesting question whether this applies to the present case as well, of which we do not know the answer.}
\begin{equation}
  \Pi(\vec{Y},\vec{X})\equiv \prod_{i=1}^N  \prod_{j=1}^M \frac{\Gamma_1(X_i-Y_j|\ep_1)}{\Gamma_1(-\ep_2+X_i-Y_j|\ep_1)}\label{eq:6}~.
\end{equation}
The derivation is very similar to that in the trigonometric case (we refer to \cite{Bullimore:2014awa,Koroteev:2015dja,Zenkevich:2017ylb} for the discussion in the context of 3d gauge theories). Indeed, consider the action of the rRS Hamiltonian in
$X$ variables on the kernel function~\eqref{eq:6}. Using the
recurrence relation for the $\Gamma_1$ function we get
\begin{equation}
  \label{eq:7}
  H_2^{(X)} \Pi(\vec{Y},\vec{X}) = \sum_{i=1}^N
  \prod_{j\neq i}^N \frac{X_i-X_j-\ep_2}{X_i-X_j} \prod_{k=1}^M
  \frac{X_i - Y_k}{X_i - Y_k - \epsilon_2} \ \Pi(\vec{Y},\vec{X})~.
\end{equation}
We can notice that the rational function in \eqref{eq:7} can be written as a contour integral
\begin{equation}
  \label{eq:9}
  H_2^{(X)} \Pi(\vec{Y},\vec{X}) = - \oint_{\mathcal{C}_X} \frac{\d z}{2\pi \i \epsilon_2}
  \prod_{j=1}^N \frac{z - X_j - \ep_2}{z - X_j} \prod_{k=1}^M
  \frac{z - Y_k}{z - Y_k - \epsilon_2} \ \Pi(\vec{Y},\vec{X})~,
\end{equation}
where $\mathcal{C}_X$ encircles the poles at $X_i$, $i=1,\dots,N$. Expanding the
contour we get residues at $z = Y_i$ and at $z = \infty$, namely 
\begin{align}
  \label{eq:11}
   & \oint_{\mathcal{C}_Y} \frac{\d z}{2\pi \i \epsilon_2}
  \prod_{j=1}^N \frac{z - X_j - \ep_2}{z - X_j} \prod_{k=1}^M
  \frac{z - Y_k}{z - Y_k - \epsilon_2} = \sum_{k=1}^M
  \prod_{j\neq k}^M \frac{Y_k-Y_j+\ep_2}{Y_k-Y_j} \prod_{l=1}^N
  \frac{Y_k - X_l}{Y_k - X_l + \epsilon_2}~,\\
 &  -\oint_{\mathcal{C}_{\infty}} \frac{\d z}{2\pi \i \epsilon_2}
  \prod_{j=1}^N \frac{z - X_j - \ep_2}{z - X_j} \prod_{k=1}^M
  \frac{z - Y_k}{z - Y_k - \epsilon_2} = N-M~.\label{eq:13}
\end{align}

Summing up two contributions~\eqref{eq:11} and~\eqref{eq:13} we get
\begin{multline}
  \label{eq:10}
  H_2^{(X)} \Pi(\vec{Y},\vec{X}) = \left[ N-M + \sum_{k=1}^M
    \prod_{j\neq k}^M \frac{Y_k-Y_j+\ep_2}{Y_k-Y_j} \prod_{l=1}^N
    \frac{Y_k - X_l}{Y_k - X_l + \epsilon_2} \right]
  \Pi(\vec{Y},\vec{X}) =\\
  = \left[ N-M + \sum_{k=1}^M \prod_{j\neq k}^M
    \frac{Y_k-Y_j+\ep_2}{Y_k-Y_j} e^{- \epsilon_1 \partial_{Y_k}}
  \right] \Pi(\vec{Y},\vec{X}) = \left[ N-M + (H_2^{(Y)})^{\dag}\right] \Pi(\vec{Y},\vec{X})~,
\end{multline}
where the conjugate of the rRS Hamiltonian is defined using the following scalar product on wavefunctions\footnote{Strictly speaking, we should carefully specify the class of allowed functions as we need some control over the analytic properties. However, here we only act on kernel functions which are well under control.}
\begin{equation}
  \label{eq:14}
  \langle f| g \rangle \equiv \oint \frac{\d^N X}{(2\pi \i)^N} \Delta^{(\epsilon_1, \epsilon_2)} (\vec{X}) f(\vec{X}) g(\vec{X})~,
\end{equation}
with measure determined by the $\ep$-deformed version of the Vandermonde determinant
\begin{equation}
  \label{eq:15}
  \Delta^{(\epsilon_1, \epsilon_2)} (\vec{X}) \equiv \prod_{i \neq j}^N
  \frac{\Gamma_1(X_i - X_j -\epsilon_2|\epsilon_1)}{\Gamma_1(X_i - X_j|\epsilon_1)}~.
\end{equation}

The integral representation of the multiparticle wavefunction is thus
given by
\begin{multline}
  \label{eq:16}
 \psi(\vec{X}^{(N)}|\vec{z}^{(N)}) \equiv z_N^{-\frac{\sum_{i=1}^N X_i^{(N)}}{\epsilon_1}}\times\\
 \times \oint \prod_{a=1}^{N-1}\frac{\d X^{(a)}  }{(2\pi
    \i)^{N/2}} \prod_{a=1}^{N-1} \left( \frac{z_a}{z_{a+1}} \right)^{-\frac{\sum_{i=1}^a
      X^{(a)}_i}{\epsilon_1}} \Delta^{(\epsilon_1,
    \epsilon_2)}(\vec{X}^{(a)})
  \Pi(\vec{X}^{(a)}, \vec{X}^{(a+1)}) ~.
\end{multline}
In fact, the $N$-body rRS Hamiltonian $H_2^{(X^{(N)})}$ acts on the last kernel, and using the
identity~\eqref{eq:10} it can be expressed through the conjugate
Hamiltonian acting on $X^{(N-1)}$ variables. The
$\epsilon$-deformed Vandermonde determinant can then be used to
convert the conjugate Hamiltonian back to the original one. Moving the Hamiltonian sequentially from right to left one eventually
proves that (\ref{eq:16}) is indeed an eigenfunction with eigenvalue
\begin{equation}
  \label{eq:18}
  E_2 = z_N \left( 1 + \frac{z_{N-1}}{z_N} \left( 1 +
      \frac{z_{N-2}}{z_{N-1}} (\ldots) \right) \right) = \sum_{i=1}^N z_i~.
\end{equation}

\textbf{Remark}. The eigenfunctions of the tRS model describe vortex partition functions of the 3d self-mirror $T[\text{U}(N)]$ theory \cite{Bullimore:2014awa}, the S-duality wall of 4d $\mathcal{N}=4$ SYM. In this setup, the tRS Hamiltonians and eigenvalues represent the action of bulk 't Hooft and Wilson loop operators at the interface. This theory also arises from maximal Higgsing of the 5d square theory $\text{U}(N)^{N-1}$ \cite{Zenkevich:2017ylb,Aprile:2018oau}, and it can also be seen a full monodromy defect in 5d $\mathcal{N}=1^*$ SYM \cite{Bullimore:2014awa}. Similarly, following the algebra/gauge dictionary of section \ref{sec:hWalg}, the eigenfunctions of the rRS model which we have constructed also match vortex partition functions of the 2d $\mathcal{N}=(2,2)$ theory arising from maximal Higgsing of the 4d square theory. This property has been known for a long time \cite{negut2009,Braverman:2010ei}, and the embedding of the theory as a full monodromy defect in 4d $\mathcal{N}=2^*$ SYM was further studied in \cite{Nawata:2014nca}. We also refer to \cite{Honda:2013uca} the for gauge theory interpretation of the Hamiltonians.  

We now turn to describing the spectral dual system. The dual tCMS Hamiltonian\footnote{This is the second in the hierarchy of
  CMS integrals of motion. The first Hamiltonian $\tilde H_1
  = \sum_{i=1}^N \epsilon_1 z_i \partial_{z_i}$ just counts the total homogeneous
  degree of the wavefunction in $z_i$.} 
\begin{equation}
  \label{eq:20}
  \tilde H_2 \equiv \sum_{i=1}^N  (\epsilon_1 z_i\partial_{z_i})^2 -  \sum_{i\neq j}
  \frac{\epsilon_2 \left(
      \epsilon_2 + \epsilon_1 \right)z_i z_j}{(z_i - z_j)^2}
\end{equation}
acts on the $z_i$ variables, but it is non-trivial to see directly from the given integral representation that (\ref{eq:16}) (up to a rescaling which we could not fix so far) is in fact its eigenfunction as well. In the
2-particle case, one can deduce the result from the hypergeometric differential equation
\be\label{eq:24}
\tilde H_{n=1,2} \tilde{\psi}(X_1,X_2|z_1, z_2) = (X_1^n +X_2^n)\tilde{\psi}(X_1,X_2|z_1, z_2)~,
\ee
which allows us to fix the correct rescaling factor as (essentially, the vacuum wavefunction which is usually stripped off from the very beginning)
\begin{equation}
  \label{eq:25}
  \tilde{\psi}(X_1,X_2|z_1, z_2) \equiv \frac{z_2}{z_1}
  \left( 1 - \frac{z_2}{z_1} \right)^{\frac{\epsilon_+}{\epsilon_1}} \psi(X_1,X_2|z_1, z_2)~.
\end{equation}
Another way of proving \eqref{eq:24} is to first use the
Euler transformation on the hypergeometric function in \eqref{eq:23}
\begin{equation}
  \label{eq:27}
  {}_2F_1\left[\begin{smallmatrix} \epsilon_+ ,&\epsilon_+ +X_1-X_2\\
      \ep_1,&\ep_1+X_1-X_2\end{smallmatrix} \left| \frac{z_2}{z_1}
    \right. \right] = \left( 1 - \frac{z_2}{z_1} \right)^{-
    2\frac{\epsilon_2}{\epsilon_1}-1} {}_2F_1\left[\begin{smallmatrix} - \epsilon_2,& - \epsilon_2+X_1-X_2\\
      \ep_1,&\ep_1+X_1-X_2\end{smallmatrix} \left| \frac{z_2}{z_1}
    \right. \right]~,
\end{equation}
and then use  the Euler-Selberg integral representation
\begin{multline}
  \label{eq:26}
  \tilde{\psi}(X_1, X_2| z_1, z_2) \sim
  (z_1 z_2)^{-\frac{X_1}{\epsilon_1}}
  \frac{z_2}{z_1} ( z_1 - z_2
  )^{-\frac{\epsilon_2}{\epsilon_1}} \frac{\e^{\frac{\i\pi}{\ep_1}(X_2-X_1)}}{\sin\frac{\pi}{\ep_1}(X_2-X_1)}
  \frac{\Gamma_1 \left( X_2 - X_1 + \epsilon_+ | \epsilon_1 \right)}{\Gamma_1 \left( X_2 - X_1 - \epsilon_2 | \epsilon_1 \right)}\times\\
  \times \oint_{\gamma_1} \d x\, x^{\frac{X_1 - X_2 - \epsilon_+ }{\epsilon_1}} (z_2 - x)^{\frac{
      \epsilon_2}{\epsilon_1}} \left( z_1 - x
  \right)^{\frac{\epsilon_2}{\epsilon_1}}~,
\end{multline}
where $\gamma_1$ is the Pochhammer contour wrapping two times around
$0$ and $z_1$. In general, the appropriately rescaled rRS eigenfunction is also an eigenfunction of the  tCMS, with eigenvalues
\begin{equation}
  \label{eq:19}
  \tilde E_n = \sum_{i=1}^N X_i^n.
\end{equation}
The dual integral representation can be written recursively using the kernel function 
\begin{equation}
  \label{eq:22}
  \text{P}(\vec{z}, \vec{w}) \equiv \prod_{i=1}^N \prod_{j=1}^M\frac{1}{(z_i - w_j)^{-\frac{\epsilon_2}{\epsilon_1}}}~,
\end{equation}
and the corresponding Euler-Selberg representation reads as
\begin{multline}
  \label{eq:21}
  \tilde\psi(\vec{X}^{(N)}|\vec{z}^{(N)}) 
  \sim \left( \prod_{i=1}^N
    z_i\right)^{-\frac{X_1}{\epsilon_1}}\times\\ \times\oint\prod_{a=1}^{N-1} \frac{\d \ln z^{(a)} }{(2\pi
    \i)^{N/2}} \prod_{a=1}^{N-1} \left( \prod_{i=1}^a
    z_i^{(a)} \right)^{\frac{ X_a - X_{a+1} - \epsilon_+}{\epsilon_1}} \Delta(\vec{z}^{(a)})^{-2\frac{\epsilon_2}{\epsilon_1}}\  \text{P}(\vec{z}^{(a)}, \vec{z}^{(a+1)})~,
\end{multline}
where $\Delta(\vec{z}) \equiv \prod_{i<j} (z_i - z_j)$ is the standard
Vandermonde determinant. 

From the last expression, it is manifest that the polynomial eigenfunctions, a.k.a. Jack polynomials, can be obtained by putting the rRS positions $\vec X^{(N)}$ on a lattice. However, this can also be seen in the Mellin-Barnes representation which is adapted to the rRS models instead. For instance, in the 2-body case, the representation~\eqref{eq:23} immediately tells us that if $X_1 - X_2 +\epsilon_+ = -n \epsilon_1 $ with $n \in \mathbb{Z}_{\geq 0}$, then
the hypergeometric series truncates and becomes a polynomial in
$z_2/z_1$. The first few Jack polynomials explicitly read 
\begin{align}
  \label{eq:2}
  \psi(X_1, X_1 + \epsilon_1 + \epsilon_2) &\sim 1,\\
  \psi(X_1, X_1 + 2\epsilon_1 + \epsilon_2) &\sim z_1 + z_2 = p_1,\\
  \psi(X_1, X_1 + 3\epsilon_1 + \epsilon_2) &\sim z_1^2 + z_2^2 +
  \frac{2(\epsilon_1 + \epsilon_2)}{(2 \epsilon_1 + \epsilon_2)} z_1
  z_2 =\frac{\ep_1}{2\ep_1+\ep_2}\left( p_2 + \frac{\epsilon_1 + \epsilon_2}{\epsilon_1} p_1^2\right)~,\\
  \cdots
\end{align}
where $p_n = z_1^n + z_2^n$ denotes the power sum basis.

Before concluding this section, let us point out that there is a further natural limit one can take in the rRS model, which from the gauge theory perspective corresponds to a \textit{semi-classical limit} removing the disk equivariant parameter. This is the \textit{non-relativistic} limit $\ep_1\to 0$ with $\beta\equiv -\ep_2/\ep_1$ fixed\footnote{From a 2d CFT perspective, this is the \textit{heavy-charge-limit} as $X/\ep_1\to \infty$ or the \text{classical limit} as $c\to \infty$.} 
\be
H_2\to \sum_i\partial_{X_i}^2+\sum_{i\neq j}\frac{\beta}{X_i-X_j}\partial_{X_i}~,
\ee
at quadratic order in $\ep_1$.  Upon taking the limit, the $\ep$-deformed Vandermonde determinant coming from the 2-point function of the $\text{W}_{\ep_1,\ep_2}$ screening currents reduces to its $\beta$-deformed version
\be
\Delta^{(\ep_1,\ep_2)}(\vec X)\to \prod_{i<j}(X_i-X_j)^{2\beta}~,
\ee
and similarly for the generalized kernel (potential terms)
\be
\prod_{i,f}\frac{\Gamma_1(X_i-Y_f|\ep_1)}{\Gamma_1(X_i-Y_f-\ep_2\alpha_f|\ep_1)}\to \prod_{i,f}(X_i-Y_f)^{-\beta\alpha_f}~.
\ee
In particular, this means that in this limit the $\text{W}_{\ep_1,\ep_2}$ free boson correlators collapse to ordinary 2d CFT Dotsenko-Fateev matrix models, and $\text{W}_{\ep_1,\ep_2}$ algebras to ordinary $\text{W}$ algebras. In the example discussed before, the (dual) positions of the matrix model correspond to the coordinates of the rRS model, while all the (dual) momenta are set to be degenerate (i.e. fixed by $\beta$). The non-relativistic rRS Hamiltonian is also equivalent to the rCMS model upon stripping off the vacuum wavefunction. This can be directly seen by performing this similarity transformation on the tCMS Hamiltonian
\be
\tilde H_2\simeq \sum_{i=1}^N(\ep_1 z_i\partial_{z_i})^2-\ep_1\ep_2\sum_{i\neq j}\frac{z_i+z_j}{z_i-z_j}z_i\partial_{z_i}~,
\ee
and then taking the rational limit by setting $z_i\equiv \exp(\ep_1 y_i)$ and keeping the constant terms in $\ep_1$
\be
\tilde H_2 \to \sum_i\partial_{y_i}^2+\sum_{i\neq j}\frac{\beta}{y_i-y_j}\partial_{y_i}~.
\ee
We see that, having now a rational/rational model, this is again self-dual. However, the non-relativistic limit has different meanings: in the rRS side, it rescales the positions by a very large parameter (i.e. heavy-charge regime), while in the tCMS side it pushes the coordinates close to the unit circle (i.e. decompactification regime).

\section{Conclusions and future directions}\label{sec:outlook}

In this paper, we have introduced a class of algebras which are naturally associated to any 4d $\mathcal{N}=2$ quiver gauge theory with unitary groups. Our construction is inspired by the parallel KP construction  regarding the 5d dimensional case and quiver $\text{W}_{q,t^{-1}}$ algebras. In fact, it essentially follows from their construction upon a suitable implementation of the dimensional reduction on the algebraic side, which we have called $\hbar$-limit. This is not the CFT-limit which reproduces the well-known AGT relation, rather our approach can be thought as its spectral dual (in the sense of integrable systems) and the resulting $\text{W}_{\ep_1,\ep_2}$ algebra are dual to ordinary $\text{W}$ algebra, hence sharing the same conformal blocks. Our construction has the advantage of being directly related to instanton calculus, while such connection is not manifest in the AGT framework. This is non-trivial interplay is partially due to the lack of manifest and simple $\mathbb{SL}(2,\mathbb{Z})$ covariance at various levels when reducing from five to four dimensions. At the level of associated integrable systems, this is due to the lack of spectral-self-duality, which is restored only after taking a semi-classical limit. 

There are a number of open questions and further directions to explore. Probably, the most urgent one would be a clearer understanding of $\text{W}_{\ep_1,\ep_2}$ algebras from the 2d CFT perspective. Intuitively, we may expect that they govern the integrability or recurrence relations of correlators by playing an analogous role in weight space as $\text{W}$ algebras do on the world-sheet. The precise mapping between the continuous free boson formalism and the discrete one might be crucial for this purpose and for establishing direct connections with previous and recent works concerning the affine Yangian and $\text{W}/\text{W}_{1+\infty}$ algebras \cite{Gaiotto:2017euk,Prochazka:2017qum}. Also, this would probably provide us with a more solid mathematical characterization of our algebras. Moreover, $\text{W}_{\ep_1,\ep_2}$ algebras might be a helpful machinery in trying to solve Toda theories through gauge or topological string theory techniques \cite{Kozcaz:2010af,Mitev:2014isa,Isachenkov:2014eya,Coman:2019eex}. 

Another very interesting direction to consider is what our construction has to say about the WZW/Liouville correspondence \cite{Ribault:2005wp,Hikida:2007tq}. According to AGT \cite{Alday:2010vg,Kozcaz:2010yp}, this relation appears when considering different types of codim-2 defects which can be engineered by either coupling d.o.f. to the bulk or as monodromy defects \cite{Gukov:2006jk,Braverman:2004vv,Braverman:2004cr}. The equivalence of the conformal blocks (through a generalized Fourier transform) is then a manifestation of an IR duality between the two gauge theory descriptions \cite{Frenkel:2015rda}. In a particular example involving the 5d $\mathcal{N}=1^*$ theory coupled to codim-2 defects, this has been related to Witten S-duality action \cite{Witten:2003ya}. Notably, this type of duality has also been related to various aspects of the (quantum) geometric Langlands correspondence \cite{Kapustin:2006pk}. In order to understand why we expect our work to be related to these topics, it is useful to recall that recently a (quantum) $q$-geometric Langlands correspondence has been put forward \cite{Aganagic:2017smx,Elliott:2018yqm}. A convenient way to think about it is as a correspondence between $q$-affine (off the critical level) and $\text{W}_{q,t^{-1}}$ conformal blocks for Langlands dual groups, which is strongly reminiscent of the WZW/Liouville setup. A key role from this perspective is played by the elliptic stable envelopes \cite{Aganagic:2016jmx}, which provide (part of) the connection between the two sides. Since these objects naturally arise when studying certain aspects of 3d mirror symmetry \cite{Rimanyi:2019zyi}, in turn closely related to spectral duality, we naturally expect that $\text{W}/\text{W}_{\ep_1,\ep_2}$ duality will fit into another intermediate layer between geometric Langlands and its $q$-deformation. In order to fully explore this possibility, the inclusion of diverse surface defects and non-simply laced algebras in our setup would be desirable \cite{Kimura:2017hez,Haouzi:2017vec}. 

Finally, another direction worth to be explored is whether T-duality of the little string can offer another (dual) derivation of the AGT relations, along the lines of \cite{Cordova:2016cmu} but now starting from the low energy description provided by 6d SYM. This may be achieved by exploiting the ideas developed in \cite{Costello:2018txb}. 

\acknowledgments

The authors thanks S. Pasquetti, E. Pomoni,  V. Schomerus and J. Teschner for valuable discussions. FN thanks CIRM for the hospitality during the BPS/CFT workshop where part of this work was
done. YZ acknowledges the
hospitality of DESY theory department where part of this work was
done. FN is supported by DESY theory group. YZ is partially supported by RFBR grants 19-02-00815,
18-51-05015-Arm, 19-51-50008-YaF, and 19-51-18006-Bolg.

\appendix

\section{Special functions}\label{sec:appFun}

In this appendix, we present a summary of the special functions used in the main text and some of their properties. For a full account we refer to \cite{Narukawa:2003}.

\textbf{Gamma function: integral definition and properties}.

\be
\Gamma_1(X|\omega)\equiv 
\frac{\e^{B_{11}(X|\omega)\ln\omega}}{\sqrt{2\pi}} \ \Gamma(X/\omega)~,
\ee
\be
B_{11}(X|\omega)\equiv  \frac{X}{\omega}-\frac{1}{2}=-\oint_{0}\frac{\d k}{2\pi\i }\frac{\e^{- k X}}{k(1-\e^{-\omega 
k})}~.
\ee
For ${\rm Re}(X)>0$, ${\rm Re}(\omega)>0$ we have the integral representation
\begin{align}\label{eq:Gamma1int}
\ln \Gamma_1(X|\omega)=& \  -\gamma B_{11}(X |\omega)+\oint_{\mathcal{C}}\frac{\d k}{2\pi\i }\ln(-k)\frac{\e^{- k X}}{k(1-\e^{-\omega 
k})}=\nn\\
=& \ -\gamma B_{11}(X |\omega)+\oint_{\mathcal{C}}\frac{\d k}{2\pi\i k}\ln(-k)\frac{\e^{- k (X-\omega/2)}}{2\sinh(\omega 
k/2)}~,
\end{align}
Therefore
\be
\ln\Gamma(X/\omega)=\oint_{\mathcal{C}}\frac{\d k}{2\pi\i }\ln(-k)\frac{\e^{- k X}}{k(1-\e^{-\omega k})}-(\gamma+\ln\omega)B_{11}(X|\omega)+\ln\frac{1}{\sqrt{2\pi}}~.
\ee
It is useful to note that
\be
\Gamma_1(X|\omega)\Gamma_1(\omega-X|\omega)=\frac{1}{S_1(X|\omega)}\equiv \frac{1}{2\sin(\pi X/\omega)}~.
\ee

\textbf{Gamma function: series definition}.

Let us consider the multiple $\zeta$-function, which is the meromorphic analytic continuation to the complex $s$-plane of the series
\be
\zeta(s;X|\vec\omega)\equiv \sum_{\vec k\in \mathbb{Z}^r_{\geq 0}}(\vec k \cdot \vec\omega+X)^{-s}~.
\ee
The Hurwitz $\zeta$-function is $\zeta(s;X|1)$, the Riemann $\zeta$-function is $\zeta(s;1|1)$. The series is convergent for $\text{Re}(s)>r$ and generic complex $\vec\omega$ all on the same side of a line through the origin. Since it is holomorphic at $s=0$, one can define the multiple Gamma function through 
\be
\Gamma_r(X|\vec\omega)\equiv \exp\left(\zeta'(s;X|\vec\omega)|_{s=0}\right)~,
\ee
where the derivative is w.r.t. the first argument. From this definition, we also get a formal series representation of the Log Gamma function, namely
\be
\ln\Gamma_1(a+X|\omega)=\zeta'(0;a|\omega)-X\sum_{k\geq 0}(k\omega+a)^{-1}+\sum_{j\geq 2}\frac{(-1)^j}{j}\zeta(j;a|\omega)X^j~.
\ee
The linear term in $X$ is divergent and needs to be interpreted as 
\be\label{eq:zetaone}
\lim_{s\to 0}\frac{\zeta(1+s;a|\omega)+\zeta(1-s;a|\omega)}{2}=-\omega^{-1}\psi_0(a/\omega)~,
\ee
where $\psi_0(X)=\Gamma'(X)/\Gamma(X)$ is the digamma function. Therefore we get the series expansion
\be\label{eq:GammaZeta}
\ln\Gamma_1(a+X|\omega)=\sum_{j\geq 0}\hat\zeta(j;a|\omega)\frac{X^j}{j!}~,
\ee
where we defined
\be
\hat\zeta(j;a|\omega)\equiv \left\{\begin{array}{ll}(-1)^j\zeta(j;a|\omega)\Gamma(j)& j>1\\
\omega^{-1}\psi_0(a/\omega)&j=1\\
\zeta'(0;a|\omega)&j=0\end{array}\right.~,
\ee 
A useful identity is 
\be\label{eq:zetagamma}
\zeta(s;a|\omega)\Gamma(s)=\int_0^\infty \frac{\d k}{k}\, \frac{k^s\, \e^{-k (a-\omega/2)}}{2\sinh(\omega k/2)}~,
\ee
which is convergent for $\text{Re}(s)>1$, $\text{Re}(a-\omega/2)>0$. For $s\to 1$ we may use (\ref{eq:zetaone}) to define the l.h.s., for $s\to 0$ we may regularize the l.h.s. by defining it as $\zeta'(0;a|\omega)$.

\textbf{$\Gamma_q$ function}. 

The $q$-Pochhammer symbol can be defined by the infinite product
\be
(x;q)_\infty\equiv \prod_{k=0}^\infty(1-q^k x)~,
\ee
for $|q|<1$ and by analytic continuation otherwise. The $\Gamma_q$ function is defined as
\be
\Gamma_q(X)\equiv (1-q)^{1-X}\frac{(q;q)_\infty}{(q^X;q)_\infty}~,
\ee
and it has the undeformed limit
\be\label{eq:gammalimit}
\lim_{q\to 1}\Gamma_q(X)=\Gamma(X)~.
\ee

\section{Continuous vs. discrete basis}\label{sec:cont-discrete}

Suppose we want to reproduce the 2-point function (\ref{eq:4dSS}) for $\ep$-Virasoro
\begin{align}
\mb{S}(X)\mb{S}(Y)=& \ :\mb{S}(X)\mb{S}(Y): \,\frac{\Gamma_1(X-Y-\ep_2|\ep_1)\Gamma_1(X-Y+\ep_1|\ep_1)}{\Gamma_1(X-Y|\ep_1)\Gamma_1(X-Y+\ep_+|\ep_1)}\, \e^{2\gamma\beta}~,
\end{align}
by using a more conventional discrete boson(s) instead of the continuous one as we considered in the main text. By looking at (\ref{eq:GammaZeta}), we may employ a field  $\mb{a}(X)$ defined by  
\begin{align}\label{eq:gbos}
\mb{a}(X)\equiv & \ \sum_{j\geq 0}(\mb{a}^+_j+(-1)^{j}\mb{a}^-_j)\frac{X^j}{j!}~,\nn\\
 [\mb{a}^+_j,\mb{a}^-_\ell]\equiv & \ \frac{1}{k! \ell!}\Big(\hat\zeta(j+\ell;-\ep_2+\i0)-\hat\zeta(j+\ell;\ep_++\i0)+\hat\zeta(j+\ell;\ep_1+\i0)-\hat\zeta(j+\ell;\i0)\Big)~,
\end{align}
with the normal ordering prescription of pushing $\mb{a}^\mp_j$ to the left/right. The $\i0$ notation denotes a regularization which we remove after computing the normal ordering functions. 

In the main text, we have been using a continuous boson to get the very same 2-point function
\be
\mb{s}(X)=\dashint \d k\, \frac{\mb{s}(k)\, \e^{ -k X}}{2\sinh(\ep_1 k/2)}~,
\ee
together with a prescription to deal with the normal ordering of the modes around $k=0$. Let us explore how the two approaches may be related. We may be tempted to identify
\be
\mb{a}_{j+}\simeq (-1)^j\int_{>0}\d k\, k^j\, \frac{\s(k)}{2\sinh(\ep_1 k/2)}~,\quad \mb{a}_{j-}\simeq -\int_{>0}\d k\, k^j\, \frac{\s(-k)}{2\sinh(\ep_1 k/2)}~,
\ee
at least for $j>0$. This is consistent with the resulting commutator
\be
[\mb{a}_{j+},\mb{a}_{\ell-}]\simeq (-1)^j\int_0^\infty\frac{\d k}{k} \, k^{j+\ell}\, \frac{\sinh(\ep_2 k/2)\ 2\cosh(\ep_+ k/2)}{\sinh(\ep_1 k/2)}~
\ee
thanks to (\ref{eq:zetagamma}).
For $j+\ell=0,1$, the identification is more problematic due to divergencies which needs to be interpreted. These divergences are expected from the small $k$ expansion of the continuous boson, where the dangerous terms indeed arise at linear order in spectral parameter. Let us also observe that we have not considered the zero modes properly here, which can indeed mix with the $j+\ell=0,1$ terms. Hopefully, a proper treatment of the zero modes will establish an exact correspondence between the formalism employing the continuous boson and a more traditional approach employing discrete modes.

\section{The generating and screening currents}\label{app:screenProp}

In this appendix, we explicitly show that the $\ep$-Virasoro generating a screening currents (\ref{eq:ThVirGen}), (\ref{eq:epScreen}) fulfill the relation (\ref{eq:commutantS}).  We first compute
\be
\mb{Y}(Z+\i 0)\mb{S}(X)= \ :\mb{Y}(Z)\mb{S}(X):\ \frac{X-Z-\ep_2/2}{X-Z+\ep_2/2-\i 0}~,  
\ee
and similarly
\be
\mb{S}(X+\i 0)\mb{Y}(Z)= \ :\mb{Y}(Z)\mb{S}(X):\ \frac{X-Z-\ep_2/2}{X-Z+\ep_2/2+\i 0}~.
\ee
We recall that  
\be
\frac{1}{X\pm\i0}=\textrm{p.v.}\frac{1}{X}\mp\i\pi\delta(X) \ , 
\ee
so that we can write
\be
\Big[\mb{Y}(Z),\mb{S}(X)\Big]=-2\pi\i \ep_2\ \delta(X-Z+\ep_2/2)\ :\mb{Y}(Z)
\mb{S}(X): ~,
\ee
and similarly
\be
\Big[\mb{Y}(Z)^{-1},\mb{S}(X)\Big]=2\pi\i \ep_2\ \delta(X-Z-\ep_2/2)\ :\mb{Y}
(Z)^{-1}\mb{S}(X):~.
\ee
Since we can identify $\s(k)=C(k)\mb{y}(k)$, we finally get 
\be
\Big[\mb{T}(Z),\mb{S}(X)\Big]=\frac{\d_{\ep_1}}{\d_{\ep_1}X}\Big(\delta(X-Z)\mb{O}(X)\Big)~,
\ee
where
\be
\mb{O}(X) = 2\pi\i \ep_1\ep_2 :\mb{Y}(X+\ep_+/2)\mb{S}(X+\ep_1/2):\ = \ \ep_1\ep_2 :\mb{Y}(X -\ep_+/
2)^{-1}\mb{S}(X-\ep_1/2):~. 
\ee

\bibliographystyle{utphys}

\providecommand{\href}[2]{#2}\begingroup\raggedright\endgroup

\end{document}